\documentclass[twocolumn, english, prd,superscriptaddress,
aps, showpacs, showkeys, amsmath, amssymb, floatfix, nofootinbib]{revtex4-1}

\pdfoutput=1
\usepackage[pdftex]{graphicx}
\usepackage[pdftex]{color}
\usepackage[colorlinks,bookmarks]{hyperref}
\usepackage{multirow}
\definecolor{linkblue}{rgb}{0,0,0.8}
\definecolor{linkgreen}{rgb}{0,0.5,0}
\hypersetup{linkcolor=linkblue, citecolor=linkgreen, urlcolor=linkblue}

\usepackage[T1]{fontenc}
\usepackage[utf8]{inputenc}
\usepackage{lmodern}
\usepackage{babel}
\usepackage{bm}
\usepackage{esint}
\usepackage{verbatim}
\usepackage[us]{datetime}
\usepackage{booktabs}
\usepackage{enumerate}
\usepackage{graphics}
\usepackage{float}
\usepackage{amsmath}
\usepackage{amsfonts}
\usepackage{amssymb}
\usepackage{tikz}
\usepackage[normalem]{ulem}

\usepackage{array}
\newcolumntype{P}[1]{>{\centering\arraybackslash}p{#1}}
\newcolumntype{M}[1]{>{\centering\arraybackslash}m{#1}}

\def\L{\mathcal{L}}
\def\E{\mathcal{E}}
\def\P{\mathcal{P}}
\def\B{\mathcal{B}}
\def\d{{\rm d}}
\DeclareMathOperator\erf{erf}

\definecolor{vale}{rgb}{0,0.5, 1.}
\definecolor{david}{rgb}{1.,.5, 0.0}

\definecolor{co1}{rgb}{0.129412, 0.4, 0.67451}
\definecolor{co2}{rgb}{0.403922, 0.662745, 0.811765}
\definecolor{co3}{rgb}{0.819608, 0.898039, 0.941176}
\definecolor{co4}{rgb}{0.968627, 0.968627, 0.968627}
\definecolor{co5}{rgb}{0.992157, 0.858824, 0.780392}
\definecolor{co6}{rgb}{0.937255, 0.541176, 0.384314}
\definecolor{co7}{rgb}{0.698039, 0.0941176, 0.168627}

\begin{document}

\title{The impact of the cosmic variance on $H_0$ on cosmological analyses}

\author{David Camarena}
\affiliation{PPGCosmo, Universidade Federal do Espírito Santo, 29075-910, Vitória, ES, Brazil}

\author{Valerio Marra}
\affiliation{Núcleo Cosmo-ufes \& Departamento de Física, Universidade Federal do Espírito Santo, 29075-910, Vitória, ES, Brazil}

\begin{abstract}
The current $3.8 \sigma$ tension between local \cite{Riess:2018byc} and global \cite{Aghanim:2016yuo} measurements of $H_0$  cannot be fully explained by the concordance  $\Lambda$CDM model. It could be produced by unknown systematics or by physics beyond the standard model. In particular, non-standard dark energy models were shown to be able to alleviate this tension.
On the other hand, it is well known that linear perturbation theory predicts a cosmic variance on the Hubble parameter $H_0$, which leads to systematic errors on its local determination.
Here, we study how including in the likelihood the cosmic variance on $H_0$ affects statistical inference.
In particular we consider the $\gamma$CDM, $w$CDM and $\gamma w$CDM parametric extensions of the standard model, which we constrain with the latest CMB, BAO, SNe Ia, RSD and $H_0$ data.
We learn two important lessons. First, the systematic error from cosmic variance is -- independently of the model -- approximately $\sigma_{\rm cv}\approx 0.88$ km s$^{-1}$ Mpc$^{-1}$ (1.2\% $H_0^{\text{loc}}$) when considering the redshift range $0.0233 \le z \le 0.15$, which is relative to the main analysis of \cite{Riess:2018byc}, and $\sigma_{\rm cv}\approx 1.5$ km s$^{-1}$ Mpc$^{-1}$ (2.1\% $H_0^{\text{loc}}$) when considering the wider redshift range $0.01 \le z \le 0.15$.
Although $\sigma_{\rm cv}$ affects the total error budget on local $H_0$, it does not significantly alleviate the tension which remains at $\approx 3 \sigma$.
Second, cosmic variance, besides shifting the constraints, can change the results of model selection: much of the statistical advantage of non-standard models is to alleviate the now-reduced tension.
We conclude that, when constraining non-standard models it is important to include the cosmic variance on $H_0$ if one wants to use the local determination of the Hubble constant by Riess et al.~\cite{Riess:2018byc}. Doing the contrary could potentially bias the conclusions.
\end{abstract}


\maketitle

\section{Introduction}

\begin{figure}[h]
  \centering
   \includegraphics[width=8cm]{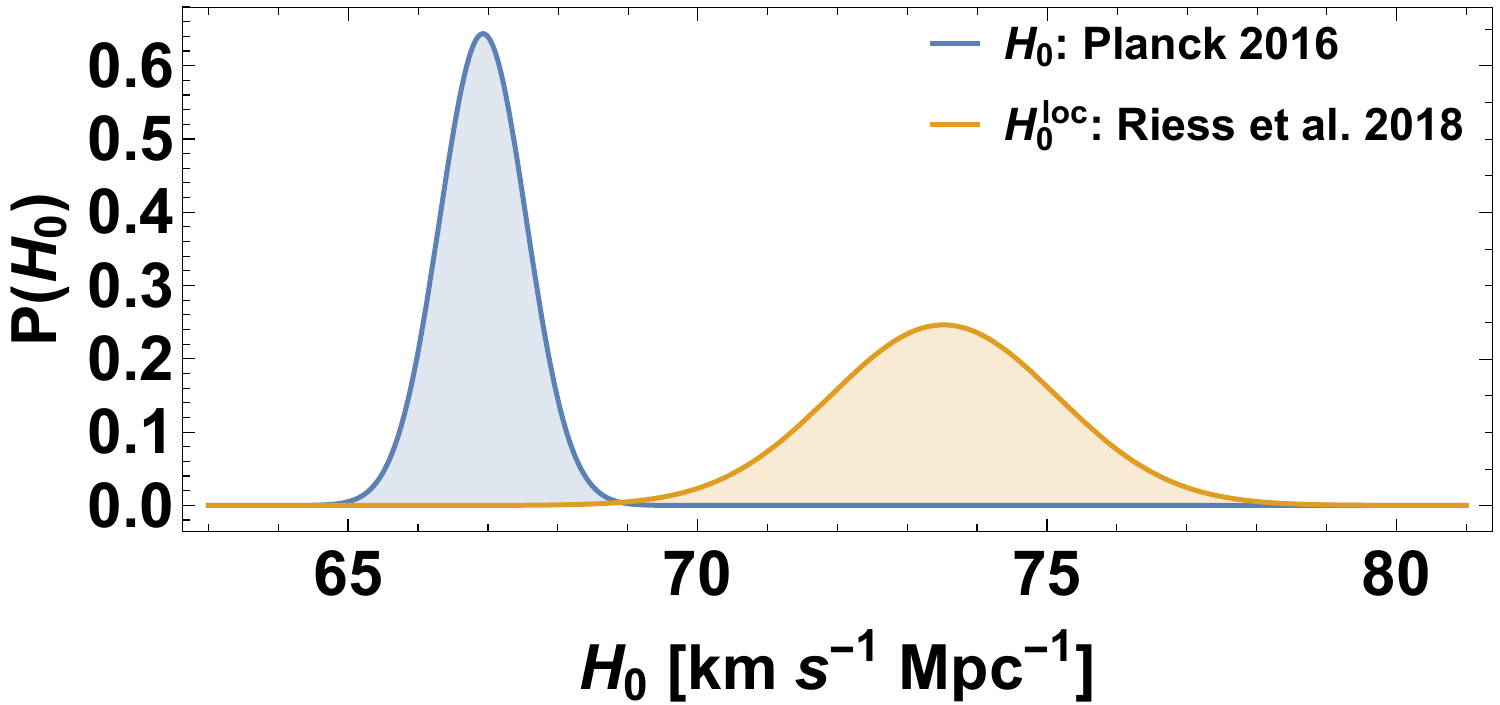}s
  \caption{Current $3.8 \sigma$ tension between local \citep[$H_0^{\text{loc}}$]{Riess:2018byc} and global \citep[$H_0$]{Aghanim:2016yuo} determinations of the Hubble constant.
  }
  \label{fig:tension}
\end{figure}

Observations of supernovae (SNe) Ia calibrated with Cepheid distances to SN Ia host galaxies \cite{Riess:2018byc} provide the value of the Hubble constant $H_0^{\text{loc}} \!\!=\!\! 73.52 \pm 1.62 \textrm{ km s}^{-1} \textrm{Mpc}^{-1}$ (hereafter $H_0^{\text{R18}}$). On the other hand, the most recent analysis of the CMB temperature fluctuations constrains the current expansion rate to $H_0 \!\!=\!\! 66.93 \pm 0.62 \textrm{ km s}^{-1} \textrm{Mpc}^{-1}$ \cite{Aghanim:2016yuo}. 
These determinations are in a tension at $ 3.8\sigma$, see Figure~\ref{fig:tension} for a visual representation.
At this moment, it is perhaps the most severe problem in the standard model, especially because it involves the well-understood physics of the CMB and the cosmological-independent analysis of the local expansion rate.

A re-assessment of the error budget of the local Hubble constant was carried out by \cite{Efstathiou:2013via,Cardona:2016ems,Zhang:2017aqn,Feeney:2017sgx} and improved near-infrared supernova measurements were considered in \cite{Dhawan:2017ywl}.
It is possible to analyze data that require unknown systematics, but this comes at the cost of obtaining degraded constraints on the cosmological parameters~\cite{Bernal:2018cxc}.
If not due to unknown systematics, it may signal physics beyond the standard model. Indeed, non-standard dark energy models were shown to be able to alleviate this tension~\cite{DiValentino:2017iww,Odderskov:2015fba,DiValentino:2016hlg,DiValentino:2017rcr,Zhao:2017urm,vanPutten:2017qte,Sola:2017znb,Qing-Guo:2016ykt}.

On the other hand, a deviation of $H_0^{\text{loc}}$ with respect to its global value $H_0$ is predicted by linear perturbation theory. This deviation, produced by the peculiar velocity field, could have non-negligible effects on determination of $H_0^{\text{loc}}$, leading to over- or underestimations of the local expansion rate. Statistically, the deviation can be quantified by a theoretical variance on $H_0^{\text{loc}}$, often dubbed cosmic variance.
A systematic error is produced by this cosmic variance, which could be important in understanding the current tension on $H_0$. 
The contribution of  cosmic variance has already been considered, in the $\Lambda$CDM context, in order to alleviate the tension, both theoretically \cite{1992AJ....103.1427T,Suto:1994kd,Shi:1995nq,Shi:1997aa,Wang:1997tp,Zehavi:1998gz,Giovanelli:1999xp,Marra:2013rba,
Ben-Dayan:2014swa} (see also \cite{Keenan:2013mfa,Redlich:2014gga,Bengaly:2015nwa,Hoscheit:2018nfl}) and through $N$-body simulations \cite{Wojtak:2013gda,Hess:2014yka,Odderskov:2014hqa,Odderskov:2015fba,Odderskov:2016rro,Wu:2017fpr,Odderskov:2017ivg} (see also \cite{Kraljic:2016acj,Hellwing:2016pdl}).
The consensus is that standard $\Lambda$CDM perturbations can alleviate the tension on $H_0$ but cannot explain it away.

Here, we study the impact of cosmic variance on statistical inference for parametric extensions of the standard model. In particular, we consider the $\gamma$CDM, $w$CDM and $\gamma w$CDM models, which we constrain with the latest CMB, BAO, SNe Ia, RSD and $H_0^{\text{loc}}$ data. 
We compare the results with and without the inclusion of the cosmic variance in the error budget of $H_0^{\text{loc}}$.
We learn two important lessons.

First, the systematic error from cosmic variance is -- independently of the model -- approximately $\sigma_{\rm cv}\approx 0.88$ km s$^{-1}$ Mpc$^{-1}$ (1.2\% $H_0^{\text{R18}}$) when considering the redshift range $0.0233 \le z \le 0.15$ and $\sigma_{\rm cv}\approx 1.5$ km s$^{-1}$ Mpc$^{-1}$ (2.1\% $H_0^{\text{R18}}$) when considering the redshift range $0.01 \le z \le 0.15$.
Although it is comparable with the uncertainty on $H_0^{\text{loc}}$  -- and so it affects the total error budget -- the tension is only reduced to 3.4$\sigma$ and 2.9$\sigma$, respectively.

Second, cosmic variance, besides shifting the constraints on the parameters correlated with $H_0$, can change the results of model selection, which we perform using the Bayes factor, the AIC \cite{Akaike1974} and BIC criteria \cite{Schwarz:1978tpv}. Indeed, much of the statistical advantage of non-standard models is to alleviate the tension which is now reduced thanks to cosmic variance. We compute the tension using the simple estimator proposed in \cite{Joudaki:2016kym}, which is a particular case of the index of inconsistency proposed in~\cite{Lin:2017ikq}.

This paper is organized as follows. In Section~\ref{gen} we review the cosmic variance on the Hubble constant predicted by linear perturbation theory and quantify the systematic error on $H_0$. In  Section~\ref{growth} we review the $\gamma$CDM, $w$CDM and $\gamma w$CDM models and discuss how a non-standard dark energy contributes to the cosmic variance on $H_0$. 
The data sets used in this work are discussed in Section~\ref{inferece}.
Statistical inference is presented in Section~\ref{inf} and was carried out using the numerical package \texttt{mBayes}, which is released together with this paper and briefly presented in Appendix~\ref{mBayes}. 
Our results are presented in Section~\ref{result} and Appendix~\ref{triplots}.
In Appendix~\ref{fisher} we list the Fisher matrices and the best-fit parameters relative to the likelihoods considered in this work.
Finally, we  conclude in Section~\ref{conclu}.
The fiducial cosmology is given in Table~\ref{tabfid}. We  assume spatial flatness.

\begin{table}[H]
\begin{center}
\begin{tabular}{m{2.5cm} m{4 cm}}
\hline
\hline
Parameter & Fiducial Value \\ \hline
$h $ & $0.6774$ \\
$\Omega_b h^2 $ & $0.0223$ \\
$\Omega_c h^2 $ & $ 0.1188$ \\
$\ln{(10^{10}\! A_s)} $ & $ 3.064$ \\
$n_s $ & $ 0.9667$ \\
$\gamma$ & $0.55$ (general relativity) \\
$w$ & $-1$ (cosmological constant)\\
\hline
\hline
\end{tabular}
\caption{Fiducial $\Lambda$CDM cosmology \cite[Planck 2015, Table 4, last column]{Ade:2015XIII}.}.
\label{tabfid}
\end{center}
\end{table}

\section{Cosmic variance on $H_0^{\text{loc}}$}\label{gen}
The peculiar velocity field, generated by the gravitational potential of the local distribution of matter, induces spatial fluctuations of the local expansion rate, $H_0^{\text{loc}}$. 
That is, an observer at $\vec{r}_i$ that measures the expansion rate $H_0^{\text{loc}}$ using $N$ objects at $\vec{r}_j$ ($j = 1,...,N$) will obtain $H_0^{\text{loc}}(\vec{r}_i) = H_0 + H_0 \, \delta_H(\vec{r}_i) $, or analogously \cite{1992AJTurner}:
\begin{equation}
\delta_H(\vec{r}_i) = \frac{H_0^{\text{loc}}(\vec{r}_i)}{H_0}-1 \ ,
\label{eq:deltaH0}
\end{equation}
where $H_0$ is the global value of the Hubble constant. 
If each object has a peculiar velocity $\vec{v}_j$, then
the deviation \eqref{eq:deltaH0} will be related to the radial component of the peculiar velocity, $\vec{v}_j \cdot \left( \vec{r}_j-\vec{r}_i\right)$. So, we can recast \eqref{eq:deltaH0} as:
\begin{equation}
\delta_H \left(\vec{r}_i \right) = \frac{1}{N} \sum_{j \neq i} \frac{\vec{v}_j}{H_0} \cdot \frac{\left(\vec{r}_j - \vec{r}_i \right)}{|\vec{r}_j - \vec{r}_i |^2} \ .
\label{eq:deltaH1}
\end{equation}
Thus, the deviation $\delta_H$ for a sphere of radius $R$, centered around $x$, is given by
\begin{equation}
\delta_{H,R} \left(\vec{x} \right) = \int d^3 y \frac{\vec{v}(y)}{H_0} \cdot \frac{\left( \vec{y} - \vec{x} \right)}{|\vec{y} - \vec{x}|^2} W ( \vec{y} - \vec{x} )  \ ,
\label{eq:deltaH2}
\end{equation}
where $W (\vec{y} - \vec{x})$ is the top-hat window function with radius $R$:
\begin{equation}
W (\vec{y} - \vec{x}) = \left\lbrace
\begin{array}{cc}
\left(4\pi R^3/3 \right)^{-1}, & |\vec{y} - \vec{x}| \leq R  \\
 \qquad 0 \qquad \quad , & \  |\vec{y} - \vec{x}| > R \ .
\end{array}
\right.
\end{equation}

Linear perturbation theory provides a relation between the peculiar velocity field and the matter density contrast $\delta$, which is
\begin{equation}
\vec{v}(y) = \frac{iH_0f}{(2\pi)^3} \int d^3 k  \frac{\hat{k} \tilde{\delta}(\vec{k}) e^{i\vec{k}\cdot \vec{y}}}{k} \ ,
\label{eq:pecuVelo-f}
\end{equation}
where $\tilde{\delta}(\vec{k})$ is the density contrast in Fourier space.
Substituting \eqref{eq:pecuVelo-f} in \eqref{eq:deltaH2} we get \cite{Shi:1995nq,Wang:1997tp,Shi:1997aa}:
\begin{equation}
\delta_{H,R} \left(\vec{x} \right) = \frac{f(z)}{(2\pi)^3} \int d^3k \, \tilde{\delta}_m \mathcal{L}(kR) e^{i\vec{k}\cdot \vec{x}}  \ ,
\label{eq:delta-CV}
\end{equation} 
where we have defined 
\begin{equation}
 \mathcal{L}(x) \equiv \frac{3}{x^3} \left( \sin{x} - \int^x_0 dy \frac{\sin{y}}{y} \right)  \ .
\label{eq:functionL}
\end{equation} 

The cosmic variance on $H_0^{\text{loc}}$ is then obtained by computing the variance of the deviation \eqref{eq:delta-CV}:
\begin{equation}
\left\langle \delta_{H}^2 \right\rangle_R = \frac{f^2(z)}{2\pi ^2 R^2} \int^{\infty}_0  dk P(k,z) [ (kR) \mathcal{L}(kR) ]^2  , 
\label{eq:finalCV}
\end{equation} 
where $P(k,z)$ is the power spectrum and the operator $\left\langle \quad  \right\rangle$ represents the ensemble (or position) average over the random fields.

\begin{figure}
\centering
\includegraphics[width=0.47\textwidth]{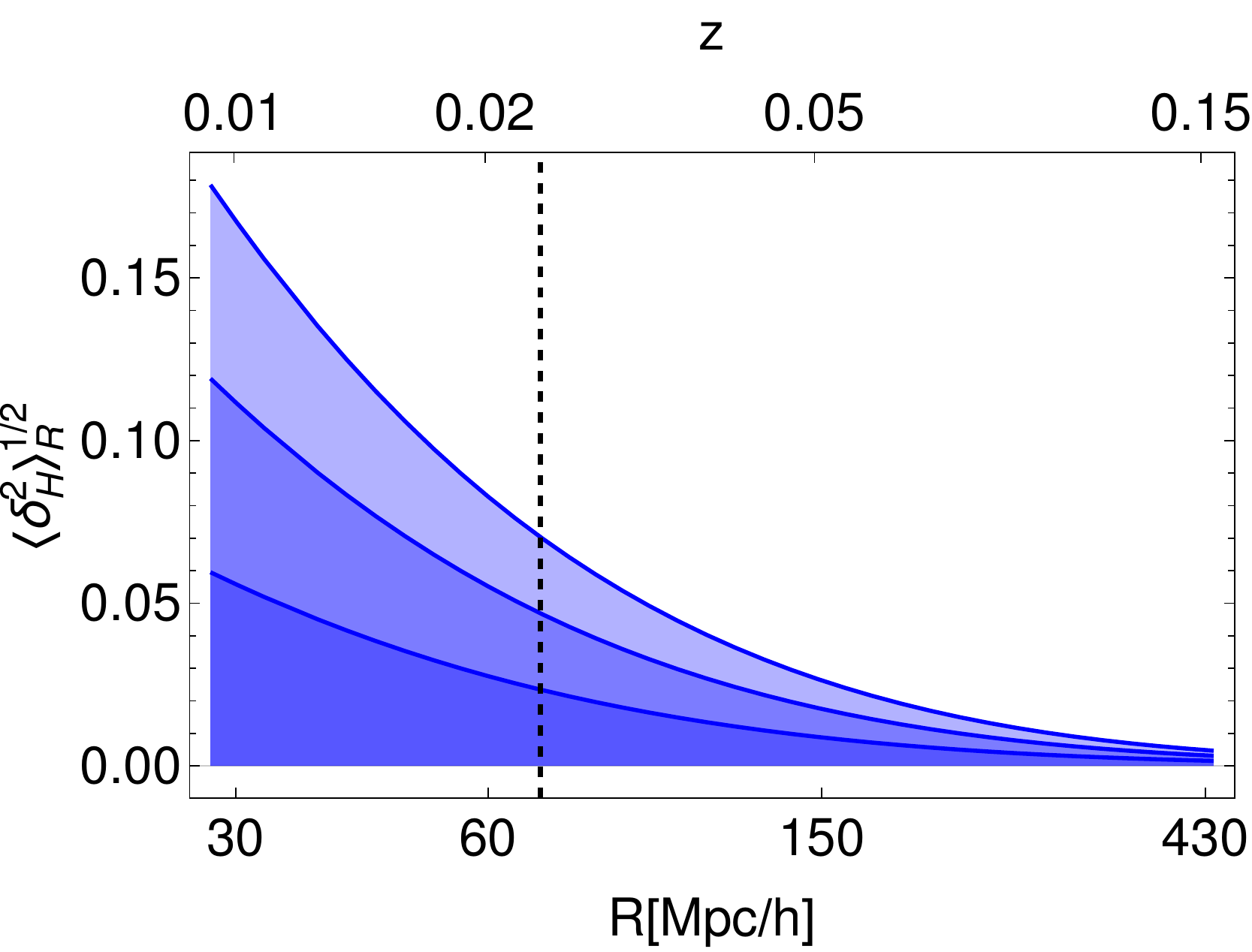}
\caption{1, 2 and 3 times the standard deviation $\left\langle \delta_{H}^2 \right\rangle_R^{1/2}$ as a function of redshift $z$ and scale $R(z)$ for the fiducial cosmology.
At larger scales there are less fluctuations on $H_0^{\text{loc}}$ because there are less matter fluctuations.
The dashed line marks the redshift $z=0.0233$, see Section~\ref{gen} for details. 
}
\label{fig:dHrz}
\end{figure}

In Figure~\ref{fig:dHrz} we plot the standard deviation $\left\langle \delta_{H}^2 \right\rangle_R^{1/2}$ in order to illustrate how it depends on the scale $R$.
At larger scales there are less fluctuations on $H_0^{\text{loc}}$ because there are less matter fluctuations. This implies that local measurements of $H_0$ have to target sources that are at a high enough redshifts so that cosmic variance is small enough and at low enough redshifts so that the measurement is still cosmology independent.
Ref.~\cite{Riess:2018byc} considers both $0.01\le z \le 0.1$ and $0.0233\le z \le 0.1$, the latter being used in the main part of the analysis as it helps to reduce cosmic variance. The redshift $z=0.0233$ is shown with a dashed line in Figure~\ref{fig:dHrz} and corresponds roughly at the scale beyond which the universe is expected to be homogeneous.

\begin{figure}
  \centering
\includegraphics[width=0.47\textwidth]{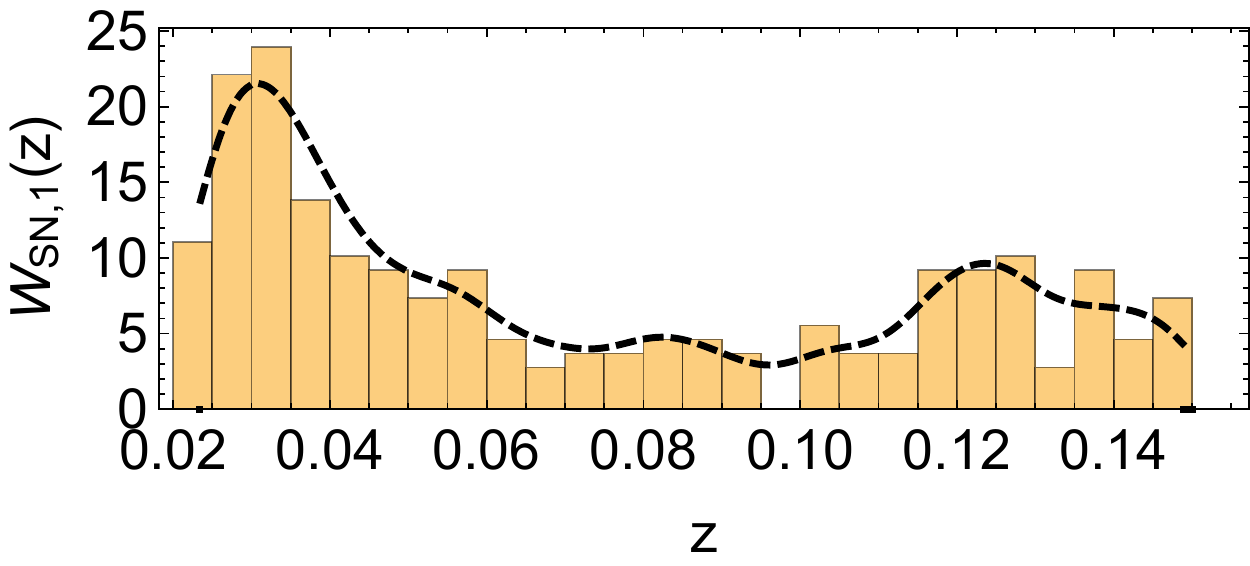}
\includegraphics[width=0.47\textwidth]{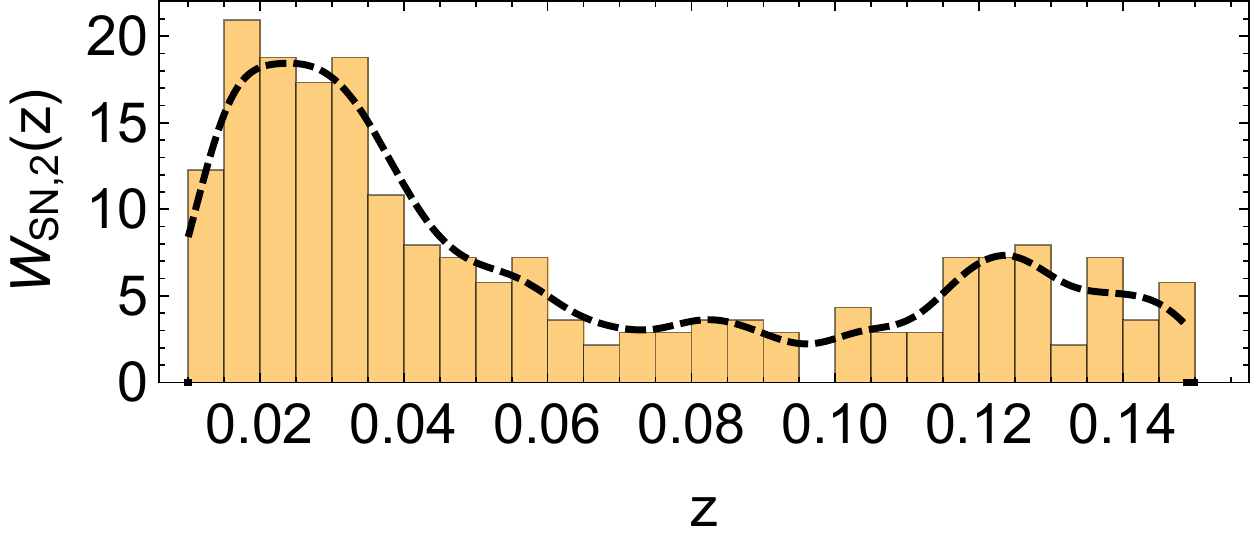}
  \caption{Distribution of SNe Ia with $0.0233\le z \le 0.15$ (top) and with $0.01\le z \le 0.15$ (bottom) which is used to obtain the redshift distribution $W_{SN}(z)$ (black dashed curve) used in equations \eqref{eq:error1} and \eqref{eq:error2}. From the Pantheon SNIa compilation~\cite{Scolnic:2017caz}.
The main analysis by Ref.~\cite{Riess:2018byc} uses the former redshift range.
  }
  \label{fig:histogram}
\end{figure}

In order to estimate the cosmic variance on \cite{Riess:2018byc} we adopt the estimator introduced in \cite{Marra:2013rba} and we consider both the redshift ranges:
\begin{align}
\sigma_{\rm cv,1} &= H_0^{\text{loc}} \left[\int_{0.0233}^{0.15} dz W_{\rm SN,1}(z)\left\langle \delta_{H}^2 \right\rangle_R  \right]^{\frac{1}{2}} , \label{eq:error1} \\
\sigma_{\rm cv,2} &= H_0^{\text{loc}} \left[\int_{0.01}^{0.15\phantom{3}} dz W_{\rm SN,2}(z)\left\langle \delta_{H}^2 \right\rangle_R  \right]^{\frac{1}{2}} , \label{eq:error2}
\end{align}
where $W_{SN}(z)$ is the normalized redshift distribution of the SNe Ia used in~\cite{Riess:2018byc}, see Figure~\ref{fig:histogram}.
This estimator neglects any effect associated to the anisotropic distribution of the supernovas.
In other words, it estimates the monopole contribution to the variance and neglects the anisotropic contributions.
As the supernovas of Figure~\ref{fig:histogram} are reasonably well distributed over the sky \cite[see][figure 1]{Deng:2018jrp}, anisotropies may remove correlations among the supernovas so that a part of the cosmic variance that is estimated with \eqref{eq:error1} is averaged away.
Also, contrary to numerical simulations, this estimator does not take into account the Milky-Way-like position of the observer. For these reasons this estimator does not reproduce results from simulations: from figure~\ref{fig:sigmaCV} one sees that \eqref{eq:error1} gives 1.2\% in the $\Lambda$CDM case while Ref.~\cite[table 1]{Wu:2017fpr} obtains 0.4\%--0.6\% depending on the methodology used.
Although less sophisticated than $N$-body-based estimators, the estimator of \eqref{eq:error1} has the advantage that can be easily computed for cosmological models for which $N$-body simulations are not available.


\section{Cosmological models}\label{growth}

As the aim of the present paper is to study the impact of cosmic variance when analyzing models beyond $\Lambda$CDM, we will now briefly summarize the parametric extensions of the standard model that will be later considered.
It is important to stress that non-standard models may feature larger cosmic variances and so affect in a non trivial way the results of statistical inference.
In particular, $\sigma_{\rm cv}$ is directly proportional to the growth rate~$f$ so that if growth rate data push towards higher growth rates one would obtain a significantly higher cosmic variance.

\subsection{$\gamma$CDM parametrization} \label{gcdm}

Within General Relativity 
the equation for the growth rate is
\begin{gather}
\frac{df}{dN} +f^2 +\left[\frac{1}{2}-\frac{3}{2} w_{de} \left(1-\Omega_m\right)\right]f - \frac{3}{2}\Omega_m \approx 0 \,. 
\label{eq:pert}
\end{gather}
There is not an analytical solution to the latter equation and the following the parametrization is commonly used:
\begin{gather}
f(z) \approx \Omega^{\gamma}_m(z) \,,
\label{eq:grate0}
\end{gather}
where $\gamma$ can be expressed as a function of $\Omega_m$ and $w$, as shown in~\cite{peebles1980large,Wang:1998gt}.
The exact $\Lambda$CDM growth rate is well described by the previous expression with $\gamma \approx 0.55$.

We will use $\gamma$ in order to study perturbative properties of a dark energy which is different from $\Lambda$. We will consider  the $\gamma=$~constant case.


\subsection{$w$CDM parametrization} \label{wacdm}

We will parametrize the equation of state of dark energy $w=p/\rho$ in order to study the background properties of a dark energy which is different from $\Lambda$.
We will consider  the $w=$~constant case.
It is important to stress that $w$ is strongly correlated with $H_0$; see the triangular plots in Appendix~\ref{triplots}.
More precisely, the high value of $H_0^{\text{R18}}$ pushes $w$ towards (somehow troubling) phantom values; in other words, the $w$CDM model can alleviate the tension between global and local determination of the Hubble constant \cite{Riess:2018byc,DiValentino:2016hlg,Qing-Guo:2016ykt}.

\subsection{$\gamma w$CDM parametrization} \label{gwcdm}

\begin{figure}
  \centering 
   \includegraphics[width=0.47\textwidth]{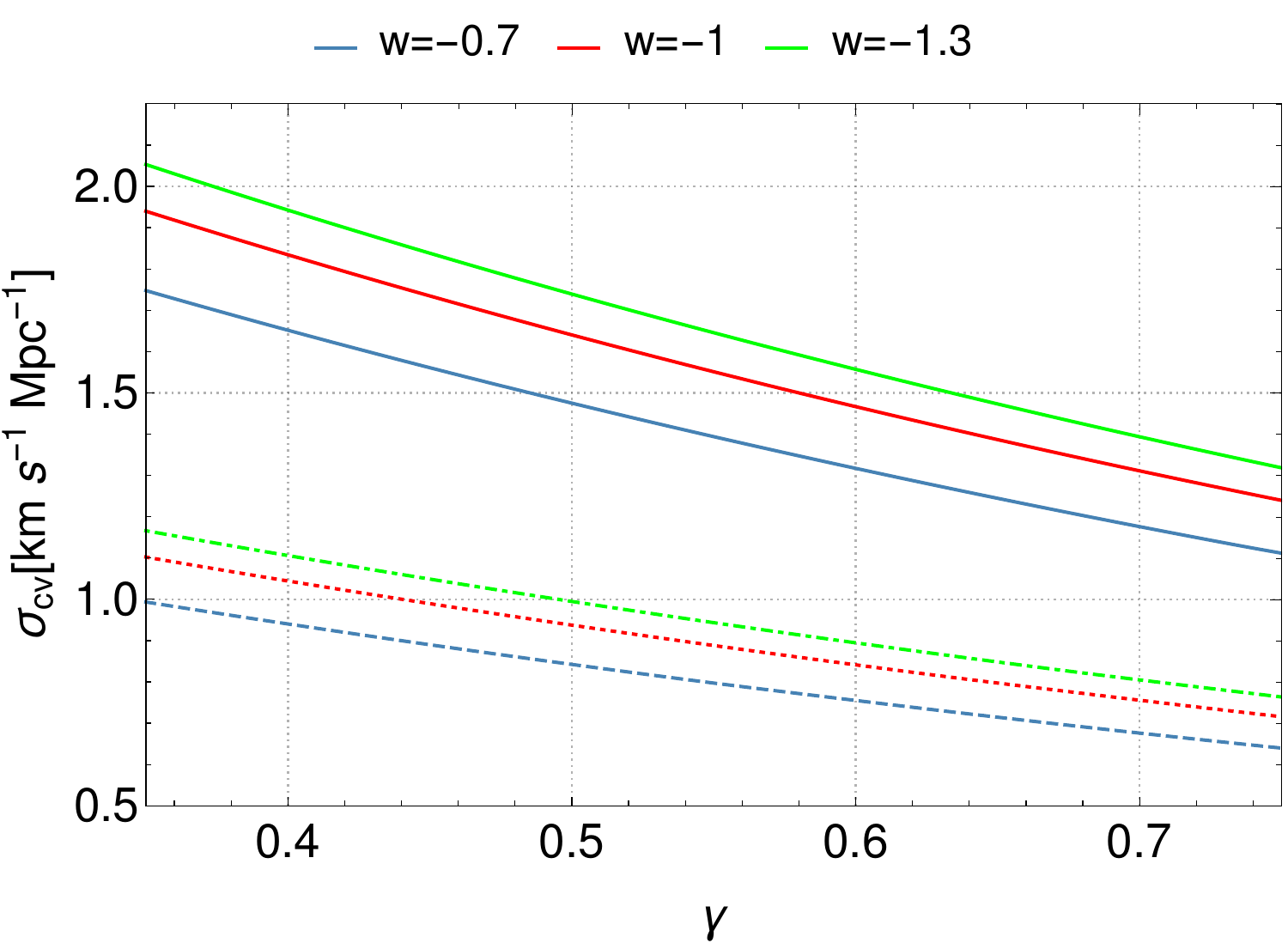}
 \caption{Cosmic-variance uncertainty $\sigma_{\rm cv}$ as a function of the growth rate parameter $\gamma$ for various values of the dark energy equation of state parameter $w$.
Dashed lines are relative to $\sigma_{\rm cv,1}$ of \eqref{eq:error1} ($0.0233\le z \le 0.15$), while solid lines to $\sigma_{\rm cv,2}$ of \eqref{eq:error2} ($0.01\le z \le 0.15$).
 }
  \label{fig:egamma}
\end{figure}

Finally, motivated by the fact that $\gamma$ and $w$ are linked by equation~\eqref{eq:pert}, we will consider the case in which both the dark energy equation of state $w$ and the growth rate parameter $\gamma$ are free to take a constant value.
Figure~\ref{fig:egamma} summarizes how the cosmic-variance uncertainty $\sigma_{\rm cv}$ depends on the growth rate parameter $\gamma$ and the dark energy equation of state parameter $w$.

\section{Data sets and likelihoods}\label{inferece}
 
In this section we present the data that we use to perform statistical inference.

\subsection{Local expansion rate}

As mentioned before, we will use the cosmology-independent  determination of the local Hubble constant by \cite{Riess:2018byc}.
Accordingly, we will build the following $\chi^2$ function:
\begin{align}
\chi^2_{H_{0, i}} & = \frac{(H_0 - H_0^{\text{R18}})^2}{\sigma^2_{{\rm loc},i}} \,.
\label{chi2H0}
\end{align}
In order to highlight the effect of the cosmic variance on statistical inference, we will consider three cases for $\sigma_{\rm loc}$ (and consequently for $\chi^2_{H_0}$):
\begin{align}
\sigma^2_{\rm loc,0} &= \sigma^2_{\rm R18} \,, \label{cv0} \\
\sigma^2_{\rm loc,1} &=\sigma^2_{\rm R18} + \sigma^2_{\rm cv,1} \label{cv1} \,, \\
\sigma^2_{\rm loc,2} &=\sigma^2_{\rm R18} + \sigma^2_{\rm cv,2} \label{cv2} \,,
\end{align}
where $\sigma_{\rm R18}\!\!=\!\! 1.62 \ \textrm{km s}^{-1} \textrm{Mpc}^{-1}$ is the uncertainty from~\cite{Riess:2018byc}.%
\footnote{We do not consider the cosmology-dependent normalization of the likelihood $\ln 2 \pi \sigma_{\rm loc}^2$ because its effect is negligible.}
As the main analysis by Ref.~\cite{Riess:2018byc} uses the redshift range $0.0233 \le z \le 0.15$, the most relevant case is the one relative to~$\sigma^2_{\rm loc,1}$.

\subsection{Cosmic Microwave Background}
The CMB is one of the most important observables in cosmology due to its well-understood linear physics, precision and sensibility to cosmological parameters. Here, we will consider the compressed CMB likelihood (Planck TT+lowP) from \cite[Table 4]{Ade:2015XIV} on the shift parameter $\mathcal{R}$, the acoustic scale $l_A$, the baryon density $\Omega_b h^2$ and the spectral index $n_s$.
The other likelihoods described in the next sections depend weakly on the latter two parameters. Therefore, in those likelihoods, we will fix $\Omega_b h^2$ and $n_s$ to their best-fit values and marginalize the CMB likelihood over $\Omega_b h^2$ and $n_s$ by eliminating the corresponding rows and columns from the covariance matrix (we adopt wide flat priors on all parameters).
$\mathcal{R}$ and $l_A$ are defined according to~\cite{Efstathiou:1998xx}:
\begin{gather}
l_A = \frac{\pi r(z_*)}{r_s(z_*)} \ , \\[2.5mm]
\mathcal{R}= \frac{\sqrt{\Omega_{m0}H_0^2}}{c} r(z_*) \ ,\label{la}
\end{gather}
where $r(z) = (1+z) d_A(z)$ is the comoving distance, $z_*$ is the redshift at decoupling and $r_s(z)$ is the sound horizon:
\begin{equation*}
r_s(z)=\frac{2}{3 k_{\rm eq}} \sqrt{\frac{6}{R_{\rm eq}}} \ln{\left(\frac{\sqrt{1+R} + \sqrt{R+R_{\rm eq}}}{1+\sqrt{R_{\rm eq}}}\right)} \ ,
\end{equation*}
with $R = \frac{3 \Omega_b (z)}{4\Omega_m (z)}$ and $R_{\rm eq}=R(z_{\rm eq})$. For $z_*$ we adopted the fit given in~\cite{Hu:1995en}.

We also consider the Gaussian likelihood on the amplitude of fluctuations $\ln{(10^{10}A_s)}$ from \cite[Table 4, first column]{Ade:2015XIII}, which, differently from $\sigma_8$, is approximately uncorrelated with respect to $\mathcal{R}$ and $l_A$. Also this likelihood is relative to the Planck TT+lowP constraints.
Consequently, we build the CMB likelihood using the following central value and Fisher matrix (or inverse covariance matrix):
\begin{align}
d_{\rm cmb}&=\{ 1.7488, \, 301.76, \, 3.089 \} \,,\\
F_{\rm cmb}&= \left(
\begin{array}{cccc}
25779 & -735.8 & 0 \\
  & 72 & 0 \\
  && 771.6
\end{array}
\right) \,,
\end{align}
and the corresponding $\chi^2$ function is:
\begin{align}
\chi^2_{\rm cmb} & = \left(d_{\rm cmb} - t_{\rm cmb}\right).F_{\rm cmb}.\left(d_{\rm cmb} - t_{\rm cmb}\right)^T 
\end{align}
where the vector  $t_{\rm cmb} = \{\mathcal{R},l_A,\ln{[10^{10}A_s]} \}$ is relative to the theoretical predictions.

\subsection{Baryonic Acoustic Oscillations}

BAO data is also of great importance for present and future cosmology, thanks again to its well-understood linear physics.
We will use BAO data from seven different surveys: 6dFGS, SDSS-LRG, BOSS-MGS, BOSS-LOWZ, WiggleZ, BOSS-CMASS, BOSS-DR12. We separate the data set in two groups, organized in Table \ref{tabBAO1} and Table \ref{tabBAO2}. 

\begin{table}
\begin{center}
\begin{tabular}{|c|c|c|c|}
\hline
Survey & $z$ & $d_z$ & $\sigma$ \\ \hline
6dFGS \cite{Beutler:2011hx} & 0.106 & 0.336 & 0.015 \\ \hline
SDSS-LRG \cite{Padmanabhan:2012hf} & 0.35 & 0.1126 & 0.0022 \\ \hline	
\end{tabular}
\caption{BAO data set in old format.}
\label{tabBAO1}
\end{center}
\end{table}
\begin{table}
\begin{center}
\begin{tabular}{|c|c|c|c|c|}
\hline
Survey & $z$ & $ \alpha^{*}\  (\text{Mpc})$ & $\sigma$ (Mpc)& $r_s^{\rm fid} \ (\text{Mpc})$ \\ \hline
BOSS-MGS \cite{Ross:2014qpa} & 0.15 & 664 & 25 & 148.69 \\ \hline
BOSS-LOWZ \cite{Anderson:2013zyy} & 0.32 & 1264 & 25 & 149.28 \\ \hline
WiggleZ \cite{Kazin:2014qga} & 0.44 & 1716 & 83 & 148.6 \\
 & 0.6 & 2221 & 101 & 148.6 \\
 & 0.73 & 2516 & 86 & 148.6 \\ \hline
BOSS-CMASS \cite{Anderson:2013zyy} & 0.57 & 2056 & 20 & 149.28 \\ \hline
BOSS-DR12 \cite{Alam:2016hwk} & 0.38 & 1477 & 16 & 147.78 \\ 
 & 0.51 & 1877 & 19 & 147.78 \\
 & 0.61 & 2140 & 22 & 147.78 \\ \hline
\end{tabular}
\caption{BAO data set in new format.}
\label{tabBAO2}
\end{center}
\end{table}

In the first case, the theoretical prediction is given by:
\begin{equation}
d_z(z)=\frac{r_s(z_d)}{D_v(z)} \,, \label{dz}
\end{equation}
so that our first $\chi^2_{\rm bao}$ function is:
\begin{equation}
\chi^2_{\rm bao,1}= \sum_i \frac{\left(d_{z,i} - d_z(z_i)\right)^2}{\sigma_i^2} \ ,
\label{chi2BAO1}
\end{equation}
where $d_{z,i}$, $\sigma_i$ and $z_i$ are given in the Table \ref{tabBAO1}. The data points are uncorrelated.

In the second case, the theoretical prediction is:
\begin{equation}
\alpha^*(z)=\frac{D_v(z)}{r_s(z_d)} r_s^{\rm fid} \,, \label{alpha}
\end{equation}
so that our second $\chi^2_{\rm bao}$ function is:
\begin{equation}
\chi^2_{\rm bao,2}= \{ \alpha^{*}_i-\alpha^{*}(z_i)\} \Sigma^{-1}_{{\rm bao},ij} \{ \alpha^{*}_j-\alpha^{*}(z_j) \} \ ,
\label{chi2BAO2}
\end{equation}
where the sum over the indices is implied and $\alpha^{*}_i$ and the corresponding $z_i$ are showed in Table~\ref{tabBAO2}.
The data points are uncorrelated, except for the WiggleZ subset.
Therefore the covariance matrix $\Sigma$ is diagonal (with variances from Table \ref{tabBAO2}) except the block relative to WiggleZ which reads:
\begin{equation}
\Sigma_{\text{WiggleZ}}= \left(
\begin{array}{ccc}
 6889 & -8961 & 21277 \\
  & 10201 & -13918 \\
  &  & 7396 \\
\end{array}
\right) \ .
\label{coWig}
\end{equation}

Note that for both $\chi^2$ functions it is necessary to compute the drag redshift $z_d$. Here, we use the fit from~\cite{Eisenstein:1997ik}. 

\subsection{Supernovae Ia}

We use the binned Pantheon SN Ia dataset~\cite[Appendix A]{Scolnic:2017caz}.
In this version of the Pantheon dataset the nuisance parameters $\alpha$, $\beta$ and $\Delta M$ are fixed at their $\Lambda$CDM best-fit values. This should not heavily bias our results as these nuisance parameters are approximately uncorrelated with respect to the cosmological parameters.

The data is given with respect to the distance modulus $\mu$ whose theoretical prediction is obtained via:
\begin{equation}
\mu(z)=5\log_{10}\frac{d_{L}(z)}{10\,\text{pc}}\,,
\end{equation}
where the luminosity distance $d_L$, for a flat universe, is given by:
\begin{equation}
d_{L}(z)=(1+z)\int_{0}^{z}\frac{d\tilde{z}}{H(\tilde{z})}\,.
\end{equation}

The $\chi^{2}$ function is then:
\begin{equation}
\chi_{\text{sne}}^{'2}= \{ \mu_{b,i} - M - \mu(z_i) \}  \Sigma^{-1}_{\text{sne},ij}  \{ \mu_{b,j} - M - \mu(z_j) \} ,
\end{equation}
where the binned distance moduli $\mu_{b,i}$, redshifts $z_i$ and covariance matrix $\Sigma$ are from the binned Pantheon catalog (considering both statistical and systematic errors). The nuisance parameter $M$ is an unknown offset sum of the supernova absolute magnitude and other possible systematics, and is completely degenerate with $\log_{10}H_{0}$.
As $M$ is not interesting as far as the present analysis is concerned, we marginalize over it right away adopting an improper prior on $M$:
\begin{equation} \label{newlike}
\mathcal{L}_{\text{sne}} = |2 \pi \Sigma_{\text{sne}}|^{-1/2} \int_{-\infty}^{+\infty} \d M e^{-\frac{1}{2}\chi_{\text{sne}}^{'2}} \,,
\end{equation}
so that one can define a new $\chi^2$ function:
\begin{equation}
\chi_{\text{sne}}^{2} \equiv -2 \ln \mathcal{L}_{\text{sne}} \,.
\end{equation}
The marginalization over $M$ can be carried out analytically. If we define the following quantities:
\begin{align}
S_{0} &=  V_{\text{1}} \cdot \Sigma_{\text{sne}}^{-1} \cdot V_{\text{1}}^{T} \,, \\
S_{1} &=  W \cdot \Sigma_{\text{sne}}^{-1} \cdot V_{\text{1}}^{T} \,, \\
S_{2} &=  W\cdot \Sigma_{\text{sne}}^{-1} \cdot W^{T} \,,
\end{align}
where $V_1$ is a row vector of unitary elements and $W_i=\mu_{b,i} - \mu(z_i)$,  one has:
\begin{equation} 
\chi_{\text{sne}}^{2}=S_{2}-\frac{S_{1}^{2}}{S_{0}} + \ln \frac{S_0}{2 \pi}+ \ln |2 \pi \Sigma_{\text{sne}}| \,, \label{chi2SNe}
\end{equation}
where the cosmology-independent normalization constants can be dropped.

\subsection{Redshift Space Distortions}

Redshift space distortion data is useful to constrain the history of structure formation and, in the coming years, will be crucial to understand the nature of dark energy. RSD data allow us to constraint the combination $f\sigma_8(z)$~\cite{Song:2008qt} and consequently the cosmic growth index $\gamma$. 
Here, we use the large RSD data compilation showed and discussed in \cite{Kazantzidis:2018rnb}. This dataset consists of 63 data points published by different surveys and is the largest compilation of $f\sigma_8 (z)$ data presented in the literature so far.
Due to overlap in the galaxy samples these data points are expected to be correlated. However, \cite{Kazantzidis:2018rnb} showed that this correlation has not a large impact on cosmological analyses. So, one can neglect correlations due to overlap and only consider the covariance matrix given for each survey. 

We can then define the following $\chi^2$:
\begin{align}
\chi'^2_{\rm rsd}&=\left\lbrace f\sigma_{8,i} - f\sigma_8(z_i) \right\rbrace \Sigma^{-1}_{\text{rsd},ij} \left\lbrace f\sigma_{8,j} - f\sigma_8(z_j) \right\rbrace  \nonumber\\
&\equiv \left\lbrace d_i - \sigma_8 t_i \right\rbrace \Sigma^{-1}_{\text{rsd},ij} \left\lbrace d_j - \sigma_8 t_j \right\rbrace \, , 
\label{chi2RSD1} 
\end{align}
where $d_i$ is the data vector, $t_i=t(z_i)$ and the theoretical prediction is given by:
\begin{equation}
f\sigma_8(z)=f(z) \, \sigma_8 \, D(z) \equiv \sigma_8 \, t(z) \,,
\end{equation}
where $\sigma_8$ is the root-mean-square mass fluctuation in spheres with radius $8h^{-1}$ Mpc at $z=0$ and $D(z)$ is the growth function normalized according to $D(0)=1$. 
The data points $z_i$ and $f\sigma_{8,i}$ are given in \cite[Table II]{Kazantzidis:2018rnb} together with the error that can be used to build the covariance matrix $\Sigma$.
We correct the prediction $t(z)$ by taking into account the fiducial model used in the analysis as explained in \cite{Kazantzidis:2018rnb,Macaulay:2013swa}.
$\Sigma$ is diagonal except for the block relative to WiggleZ which reads:
\begin{equation}
\Sigma_{\rm WiggleZ} = 10^{-3}\left(
\begin{array}{ccc}
 6.4 & 2.57 & 2.54 \\
  & 3.969 & 2.54 \\
  &  & 5.184 \\
\end{array}
\right) \, .
\label{coWig2}
\end{equation}

The $\chi^2$ function of equation~\eqref{chi2RSD1} depends on $\sigma_8$. However, RSD data were obtained assuming the $\Lambda$CDM model; in particular, it is assumed the standard initial power spectrum, which may have evolved differently for alternative theories that feature a different matter era.
Therefore, we conservatively marginalize over $\sigma_8$ as the latter is degenerate with the initial conditions of the perturbations.
This means that only the curvature of $f\sigma_8(z)$ matters and not its overall normalization.
We will not consider changes in $\alpha=\delta'_{\rm inicial}/\delta_{\rm inicial}$, that is, we assume that at high redshift the standard cosmology is valid ($\alpha=1$), see~\cite{Taddei:2014wqa,Taddei:2016iku} for a thorough discussion.

As $\sigma_8$ is not interesting as far as the present analysis is concerned, we marginalize over it right away adopting an improper flat prior on $\sigma_8>0$:
\begin{equation} \label{newlikeRSD}
\mathcal{L}_{\text{rsd}} = |2 \pi \Sigma_{\text{rsd}}|^{-1/2} \int_{0}^{+\infty} \d \sigma_8 e^{-\frac{1}{2}\chi_{\text{rsd}}^{'2}} \,,
\end{equation}
where it is worth stressing that, here, the parameter $\sigma_8$ is seen as a nuisance parameter; in particular, it is not the $\sigma_8$ relative to a cosmological model we may analyze.

Also in this case the marginalization can be carried out analytically.
Let us define the following auxiliary functions:
\begin{gather}
S_{dd} = d_i \Sigma^{-1}_{\text{rsd},ij} d_j \,, \nonumber \\
S_{dt} = d_i \Sigma^{-1}_{\text{rsd},ij} t_j  \,, \nonumber \\
S_{tt} =  t_i \Sigma^{-1}_{\text{rsd},ij} t_j    \,.
\end{gather}
We find then that:
\begin{align} 
\chi_{\text{rsd}}^{2} &\equiv -2 \ln \mathcal{L}_{\text{rsd}}=S_{dd}-\frac{S_{dt}^{2}}{S_{tt}} + \ln S_{tt}  \nonumber \\
&- 2 \ln \left(1+ \erf \frac{S_{dt}}{\sqrt{2 S_{tt}}} \right)
 + \ln |2 \pi \Sigma_{\text{rsd}}| \,, \label{chi2RSD}
\end{align}
where the cosmology-independent normalization constant can be dropped.

\section{Statistical inference}\label{inf}

\subsection{Total likelihood}

The total likelihood is given by:
\begin{align} \label{litot}
\chi^2_{{\rm tot}, i} &\equiv  -2 \ln \mathcal{L}_{\text{tot}} =  \chi^2_{H_{0, i}} + \chi^2_{\rm cmb}   \\
&+ \chi^2_{\rm bao,1} + \chi^2_{\rm bao,2} + \chi^2_{\rm sne} + \chi^2_{\rm rsd} \,. \nonumber
\end{align}
where the index $i$ labels the three cases of equations~(\ref{cv0}-\ref{cv2}).

We should point out that all data used here, excluding $H_0^{\rm R18}$, are model-dependent, i.e.~they use a fiducial $\Lambda$CDM model in their analyses. This could bias our results towards $\Lambda$CDM; yet this bias should not be important as the cosmologies we consider are parameterizations of the $\Lambda$CDM model.

\subsection{Measuring the tension}

\begin{table}
\begin{tabular}{M{2.5cm}|M{4.5cm}}
\hline
\hline
$T_{H0}$ & Qualitative  interpretation \\
\hline
< 1.4 & No significant tension \\
1.4 -- 2.2 & Weak tension \\
2.2 -- 3.1 & Moderate  tension \\
> 3.1 & Strong  tension \\
\hline 
\hline
\end{tabular}
\caption{Qualitative interpretation of the tension estimator $T_{H0}$ according to Jeffreys' scale of Table~\ref{jeffreys} as proposed in~\cite{Lin:2017ikq}.}
\label{tab:tension}
\end{table}

We adopt the following estimator%
\footnote{This estimator was used in \citep{Joudaki:2016kym} to asses the  $S_8$ $(\equiv \sigma_8 \sqrt{\Omega_m/0.3})$ tension.
More sophisticated estimators can be found in the literature. For instance, the tension $\mathcal{T}$ \citep{Verde:2013wza} or the index of inconsistency IOI \cite{Lin:2017ikq}.}
in order to quantity the discordance or tension in current determinations of~$H_0$:
\begin{equation}
T_{H0}=\frac{|H_0 - H_0^{\rm R18}|}{\sqrt{\sigma_{H0}^2 + \sigma_{\rm loc}^2}} \ ,
\label{ten}
\end{equation}
where $H_0$ and $\sigma^2_{H0}$ are mean and variance of the posterior $p(H_0)$, respectively.
In the Gaussian and weak prior 
limit the index of inconsistency, defined in \cite{Lin:2017ikq}, is IOI$= \frac{1}{2}T^2_{H_0}$. 
Thus, we can recalibrate Table~III of \cite{Lin:2017ikq} into Table~\ref{tab:tension} in order to obtain a qualitative assessment of the tension in the Hubble constant.

Using \eqref{ten} with \eqref{cv0}, that is, neglecting cosmic variance, the tension between global and local $H_0$ is about $3.8\sigma$. According to Table~\ref{tab:tension}, there is a strong tension 
(or inconsistency) between the two determinations.
If one considers the effect of the cosmic variance and uses \eqref{cv1} and \eqref{cv2}, the discordance is reduced to 3.4$\sigma$ and 2.9$\sigma$, respectively.
As the main analysis of \cite{Riess:2018byc} uses \eqref{cv1}, it seems as if cosmic variance does not have an important effect.
However, as we will see, it does have an important impact on model selection.

Hereafter, we shall compute \eqref{ten} using the $\sigma_{\rm loc,i}$ of equations (\ref{cv0}-\ref{cv2}) that is relative to the $\chi^2_{{\rm tot}, i}$ used.

\subsection{Model selection: evidence}

The natural way to perform model selection within Bayesian inference is to compare the evidences of the models via the Bayes factor.
The Bayesian evidence of a model is obtained by integrating the product of the prior $\P(T)$ and the likelihood $\L (D|T)$ over the relevant parameter space:
\begin{equation}
\E = \int \P(T) \L (D|T) \d T \,.
\end{equation}
The evidence is the normalizing factor that transforms $\P(T) \L (D|T)$ into the posterior distribution.
In the previous equations $T=\{\theta_\gamma \}$ represents the parameter vector and $D$ the dataset.
As the evidence is the likelihood of the model itself, assuming that different models have the same prior probability, one can take the ratio of the posterior probabilities of the models $i$ and $j$ and obtain the Bayes factor:
\begin{equation}
\B_{ij}= \frac{\E_i}{\E_j} \,.
\end{equation}
The above odds ratio is then interpreted qualitatively via the Jeffreys' scale \cite{jeffreys1961theory}.
Here, we will use the conservative version defined in~\cite{Trotta:2008qt}, see table~\ref{jeffreys}.

\begin{table}
\centering
\begin{tabular}{>{\centering\arraybackslash} m{1.5cm} | >{\centering\arraybackslash} m{5cm} | >{\centering\arraybackslash} m{1.5cm}}
\hline
\hline
$\ln B_{i0}$ & Strength of evidence & color code\\
\hline
$> 5$ & Strong evidence for model $i$ & \begin{tikzpicture}\filldraw[fill=co7, draw=black] (0,0) rectangle (.5,.5);\end{tikzpicture} \\ 
$[2.5, 5] $& Moderate evidence for model $i$ & \begin{tikzpicture}\filldraw[fill=co6, draw=black] (0,0) rectangle (.5,.5);\end{tikzpicture} \\ 
$[1, 2.5]$ & Weak evidence for model $i$ & \begin{tikzpicture}\filldraw[fill=co5, draw=black] (0,0) rectangle (.5,.5);\end{tikzpicture} \\ 
$[-1, 1] $& Inconclusive & \begin{tikzpicture}\filldraw[fill=co4, draw=black] (0,0) rectangle (.5,.5);\end{tikzpicture} \\ 
$[-2.5, -1]$ & Weak evidence for $\Lambda$CDM & \begin{tikzpicture}\filldraw[fill=co3, draw=black] (0,0) rectangle (.5,.5);\end{tikzpicture} \\ 
$[-5, -2.5]$ & Moderate evidence for $\Lambda$CDM & \begin{tikzpicture}\filldraw[fill=co2, draw=black] (0,0) rectangle (.5,.5);\end{tikzpicture} \\ 
$<- 5$ & Strong evidence for $\Lambda$CDM & \begin{tikzpicture}\filldraw[fill=co1, draw=black] (0,0) rectangle (.5,.5);\end{tikzpicture} \\ 
\hline
\hline
\end{tabular}
\caption{Jeffreys' scale as presented in~\cite{Trotta:2008qt}.\label{jeffreys}}
\end{table}

For the datasets and models of this work, the likelihood of \eqref{litot} can be very well approximated via the following multivariate Gaussian distribution:
\begin{equation}
\L (D|\theta_\gamma) \simeq \L_{\rm max} \, e^{- \frac{1}{2} (\theta_\alpha - \hat \theta_\alpha) L_{\alpha \beta} (\theta_\beta - \hat \theta_\beta)} \,,
\end{equation}
where $\hat \theta_\gamma$ denotes the best-fit parameters that maximize the likelihood, $\L(D|\hat \theta_\gamma)=\L_{\rm max}$, and $ L_{\alpha \beta} $ is the Fisher matrix associated to the likelihood.
As we are using wide flat (constant) priors, we can  compute analytically the evidence:
\begin{align}
\E=   \L_{\rm max} \frac{(2 \pi)^{n/2}}{|L|^{1/2}} \prod_{\alpha=1}^k \frac{1}{\Delta \theta_\alpha} \,,
\end{align}
where $k$ is the number of parameters $\theta_\gamma$, and $\Delta \theta_\gamma$ are the widths of the (possibly improper) priors.

The $\Lambda$CDM model is clearly a particular case of the $\gamma$CDM, $w$CDM and $\gamma w$CDM models considered here. Therefore, common parameters share the same priors so that the Bayes factors with respect to $\Lambda$CDM (model $j=0$) are:
\begin{align}
&\ln \B_{\gamma 0} =-\ln \Delta \gamma - \frac{1}{2} \Delta \chi^2 + \frac{1}{2} \ln \frac{|L_0|}{|L_\gamma|} + \ln \sqrt{2 \pi} \,,\\
&\ln \B_{w 0} = -\ln \Delta w - \frac{1}{2} \Delta \chi^2 + \frac{1}{2} \ln \frac{|L_0|}{|L_w|} + \ln \sqrt{2 \pi} \,,  \nonumber \\
&\ln \B_{\gamma w 0} =   -\ln \Delta \gamma -\ln \Delta w - \frac{1}{2} \Delta \chi^2 + \frac{1}{2} \ln \frac{|L_0|}{|L_{\gamma w}|} + \ln 2 \pi \,. \nonumber
\label{eq:bayesfactor}
\end{align}
Note that the common improper priors on $M$ and $\sigma_8$ cancel out when taking the ratio of the evidences.
Note also that  $\Delta \chi^2$ is only a part of the Bayes factor, and that $\ln \B<0$ supports the $\Lambda$CDM model (a positive  $\Delta \chi^2$ means that the alternative model has a worse fit as compared to $\Lambda$CDM).
The prior widths $\Delta \gamma$ and $\Delta w$ together with the ratio of the determinants of the Fisher matrices quantify the qualitative Occam's razor.
The Fisher matrices $L_{\alpha \beta}$ together with the best-fit parameters are given in Appendix~\ref{fisher}.

\subsection{Model selection: AIC and BIC}

For completeness, we consider also the Akaike information criterion (AIC) \citep{Akaike1974} and Bayesian information criterion (BIC) \citep{Schwarz:1978tpv}, which are supposed to approximate the full evidence of the previous section. They are defined according to: 
\begin{align}
\rm{AIC} &= \chi^2_{\rm min}  + 2k \ , \\
\rm{BIC} &= \chi^2_{\rm min} + k \ln{N} \ ,
\end{align}
where $N$ is the total number of data points, $k$ the number of free parameters and
\begin{equation}
\chi^2_{\rm min}\equiv -2 \ln{\mathcal{L}}_{\rm max} \,,
\end{equation}
where $\mathcal{L}_{\rm max}$ is the maximum value of the likelihood given in \eqref{litot}.
For the present analysis it is $\ln N \simeq 4.8$.
We will compute the differences $\bigtriangleup $AIC and $\bigtriangleup$BIC with respect to the standard $\Lambda$CDM model:
\begin{align}
\bigtriangleup \textrm{AIC} &=\Delta \chi^2 + 2 \Delta k \ , \\
\bigtriangleup \textrm{BIC} &= \Delta \chi^2 + (\ln N) \Delta k \ ,
\end{align}
with $\Delta k=1$ for the $\gamma$CDM and $w$CDM models and $\Delta k=2$ for the $\gamma w$CDM model.
Note that a positive value of $\bigtriangleup \textrm{AIC}$ or $\bigtriangleup \textrm{BIC}$ means a preference for $\Lambda$CDM.
Unlike $\Delta \chi^2$, the AIC and BIC criteria punish the model with a larger number of free parameters.
The values that we will obtain for the differences $\bigtriangleup $AIC and $\bigtriangleup$BIC will be interpreted according to the calibrated Jeffreys' scales 
showed in the Tables~\ref{tabAIC}%
\footnote{Note that the categories of Tables~\ref{tabAIC} do not cover the interval $[0,\infty)$. This means that these values have to be interpreted as orders of magnitudes.}
-\ref{tabBIC}.

\begin{table}
\begin{tabular}{M{2.5cm}|M{4.5cm}}
\hline
\hline
|$\bigtriangleup$AIC| & Level of empirical support for the model with the higher AIC \\
\hline
0 -- 2 & Substantial \\
4 -- 7 & Considerably less \\
>10 & Essentially none \\
\hline 
\hline
\end{tabular}
\caption{Qualitative interpretation of $\bigtriangleup$AIC according to the calibrated Jeffreys' scale \cite{Rivera:2016zzr,burnham2013model}.
}
\label{tabAIC}
\end{table}

\begin{table}
\begin{tabular}{M{2.5cm}|M{4.5cm}}
\hline
\hline
|$\bigtriangleup$BIC| & Evidence against the model with the higher BIC \\
\hline
0 -- 2 & Weak \\
2 -- 6 & Positive \\
6 -- 10 & Strong \\
>10 & Very strong \\ \hline 
\hline
\end{tabular}
\caption{Qualitative interpretation of $\bigtriangleup$BIC according to the calibrated Jeffreys' scale \cite{Rivera:2016zzr,Perez-Romero:2017njc}.}
\label{tabBIC}
\end{table}

\section{Results and discussion}\label{result}

We have performed a full Bayesian analysis of the $\Lambda$CDM, $\gamma$CDM, $w$CDM and $\gamma w$CDM models with and without considering the cosmic variance on $H_0$.
The corresponding triangular plots are shown in the Figures~\ref{fig:dH1_combo_r5}-\ref{fig:dH6_combo_r5} of Appendix~\ref{triplots}.
The plots show the strong correlation of $H_0$ with $\Omega_{m 0}$ and $w$. Therefore, any bias in the likelihood relative to $H_0$ directly translates into a bias on these parameters.
In particular, the inclusion of cosmic variance shifts the posteriors relative to $w$ towards non-phantom values.
This is shown by the $3 \sigma$ confidence levels reported in table \ref{tabRX}.
It is interesting to point out that the posterior on $H_0$ shifts towards lower values when including $\sigma_{\rm cv}$ not only because the local determination has lower statistical weight (larger error) but also because $\sigma_{\rm cv}$ depends on $\Omega_{m0}$ via the growth rate $f = \Omega_m(z)^{\gamma}$ and $\Omega_{m0}$ is inversely correlated with respect to $H_0$ (see triangular plots). Indeed, a larger $\sigma_{\rm cv}$ decreases the $\chi^2$ and can be obtained with a higher $\Omega_{m0}$ which in turns imply a lower $H_0$.
We also report the confidence levels on $\gamma$, which have a reduced constraining power because we have marginalized the posterior over the RSD normalization.
Nevertheless, the allowed values for $\gamma$ decrease when $\sigma_{\rm cv}$ is included in the analysis. This because cosmic variance is inversely proportionally to $\gamma$, see Figure~\ref{fig:egamma} (lower $\gamma$, faster growth).

\begin{table*}
\centering
\begin{tabular}{|l|c|c|c|c|c|c|c|c|c|c|c|}
\hline
\multicolumn{8}{|c|}{\bf Analysis with $\chi^2_{\rm tot,0}$ (without cosmic variance on $H_0$)} \\ \hline
Model &  3$\sigma$ c.l.~on $\gamma$ & 3$\sigma$ c.l.~on $w$ & \quad $T_{H0}$ \quad & $\Delta \chi^2$ & $\Delta$AIC & $\Delta$BIC & - \\ \hline
$\Lambda$CDM &  - & -  & $3.9$ & - & - & -& - \\ \hline
$\gamma$CDM &  $ [0.47, 0.84]$ &  - & $4.0$ & $-2.4$ & $-0.4$ & $2.3$& - \\ \hline
$w$CDM &  - &  $[-1.22,-1.03]$  & $1.6$ & $-15.8$ & $-13.8$ & $-11$& - \\ \hline
$\gamma w$CDM   & $[0.42,0.79]$ &  $[-1.22,-1.02]$ & $1.8$ & $-16.4$ & $-12.4$ & $-6.9$& - \\
\hline
\hline
\multicolumn{8}{|c|}{\bf Analysis with $\chi^2_{\rm tot,1}$ (with cosmic variance $\sigma_{\rm cv,1}$ on $H_0$)} \\ \hline
Model &  3$\sigma$ c.l.~on $\gamma$ & 3$\sigma$ c.l.~on $w$ &\quad $T_{H0}$ \quad  & $\Delta \chi^2$ & $\Delta$AIC & $\Delta$BIC &$ \chi_{\text{min},1}^2 - \chi_{\text{min},0}^2$ \\ \hline
$\Lambda$CDM &  - & -  & $3.6$ & - & - & -& $-4.1$ \\ \hline
$\gamma$CDM &  $[0.46, 0.84]$ &  - & $3.7$ & $-2$ & $0$ & $2.8$& $-3.6$ \\ \hline
$w$CDM &  - &$[-1.21,-1.02]$   & $1.7$ & $-12.8$ & $-10.8$ & $-8$&$-1.1$ \\ \hline
$\gamma w$CDM  &  $[0.42,0.79]$ &  $[-1.21,-1.01]$ & $1.8$ & $-13.5$ & $-9.5$ & $-3.9$& $-1.1$ \\
\hline
\hline
\multicolumn{8}{|c|}{\bf Analysis with $\chi^2_{\rm tot,2}$ (with cosmic variance $\sigma_{\rm cv,2}$ on $H_0$)} \\ \hline
Model &  3$\sigma$ c.l.~on $\gamma$ & 3$\sigma$ c.l.~on $w$ &\quad $T_{H0}$ \quad  & $\Delta \chi^2$ & $\Delta$AIC & $\Delta$BIC &$ \chi_{\text{min},2}^2 - \chi_{\text{min},0}^2$ \\ \hline
$\Lambda$CDM &  - & -  & $3.0$ & - & - & -& $-8.6$ \\ \hline
$\gamma$CDM &  $[0.45, 0.83]$ &  - & $3.2$ & $-1.8$ & $0.2$ & $2.9$& $-8$ \\ \hline
$w$CDM &  - &$[-1.21,-1.00]$   & $1.6$ & $-9.6$ & $-7.6$ & $-4.8$&$-2.4$ \\ \hline
$\gamma w$CDM  &  $[0.43,0.80]$ &  $[-1.20,-1.00]$ & $1.7$ & $-10.4$ & $-6.4$ & $-0.8$& $-2.6$ \\
\hline
\end{tabular}
\caption{Summary of results. See section~\ref{result} for details.}
\label{tabRX}
\end{table*} 

\begin{figure}
\centering 
\includegraphics[width=0.5\textwidth]{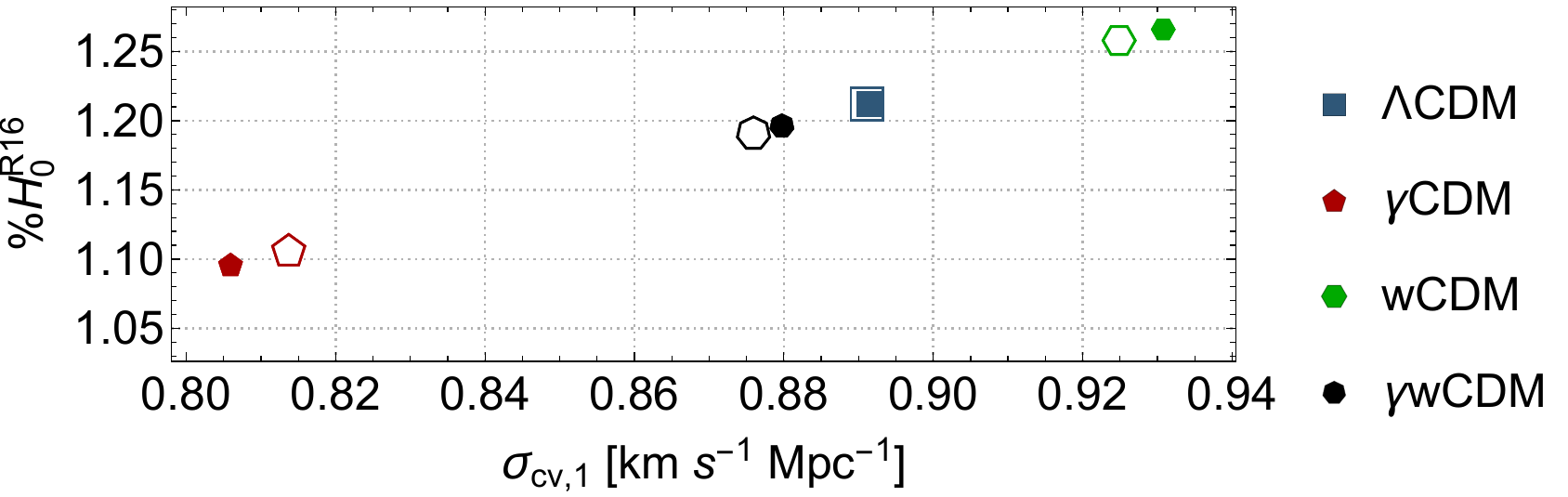}\\
\vspace{.4cm}
\includegraphics[width=0.5\textwidth]{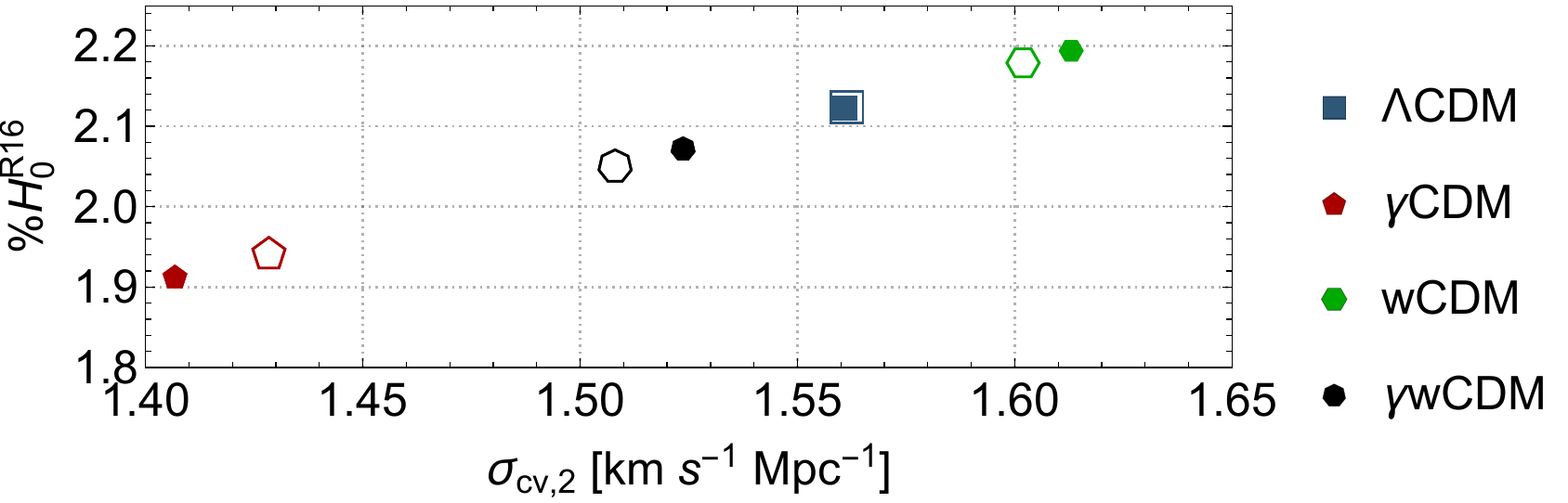}
 \caption{Cosmic variance on the Hubble constant $\sigma_{\rm cv}$ relative to the best fits of the models considered in this analysis. Filled markers refers to the analysis that uses $\chi^2_{\rm tot,0}$, while empty markers refers to $\chi^2_{\rm tot,1}$ (top panel, $0.0233\le z \le 0.15$) or $\chi^2_{\rm tot,2}$ (bottom panel, $0.01\le z \le 0.15$).}
  \label{fig:sigmaCV}
\end{figure}

In figure~\ref{fig:sigmaCV} we show the values of $\sigma_{\rm cv}$ relative to the best fits of the models considered in this analysis (given in Table~\ref{tabBFX}). Roughly, one can say that, with little variation, $\sigma_{\rm cv}\approx 0.88$ km s$^{-1}$ Mpc$^{-1}$ (1.2\% $H_0^{\text{R18}}$) when considering the redshift range $0.0233 \le z \le 0.15$ and $\sigma_{\rm cv}\approx 1.5$ km s$^{-1}$ Mpc$^{-1}$ (2.1\% $H_0^{\text{R18}}$) when considering the redshift range $0.01 \le z \le 0.15$.
This implies that one may roughly estimate the error due to cosmic variance by assuming the latter values in equation \eqref{cv1} and \eqref{cv2}, without going through the method detailed in Section~\ref{gen}.

Next we discuss model selection. First, the inclusion of the cosmic variance $\sigma_{\rm cv}$ significantly decreases the value of $\chi^2_{\rm min}$ (last column of table \ref{tabRX}). 
However, the decrease is less pronounced for the models which feature the parameter $w$. This causes the $\Delta \chi^2$ differences to decrease significantly when $\sigma_{\rm cv}$ is included (fifth column of table \ref{tabRX}).
Models with the $w$ parameter perform better because they can produce a higher $H_0$, see Table~\ref{tabBFX}; this is also shown by the fourth column of table \ref{tabRX} which shows how low the discordance on $H_0$ becomes for these models. A qualitative interpretation of the values of $T_{H0}$ is given in table~\ref{tab:tension}.
It is also worth commenting that the inclusion of $\sigma_{\rm cv}$ decreases the allowed valued of $H_0$; this is welcome since it is not trivial to accommodate a higher value of $H_0$ with the constraints from CMB, see~\cite{Evslin:2017qdn} for a discussion.

\begin{figure}
\centering
\includegraphics[width=0.47\textwidth]{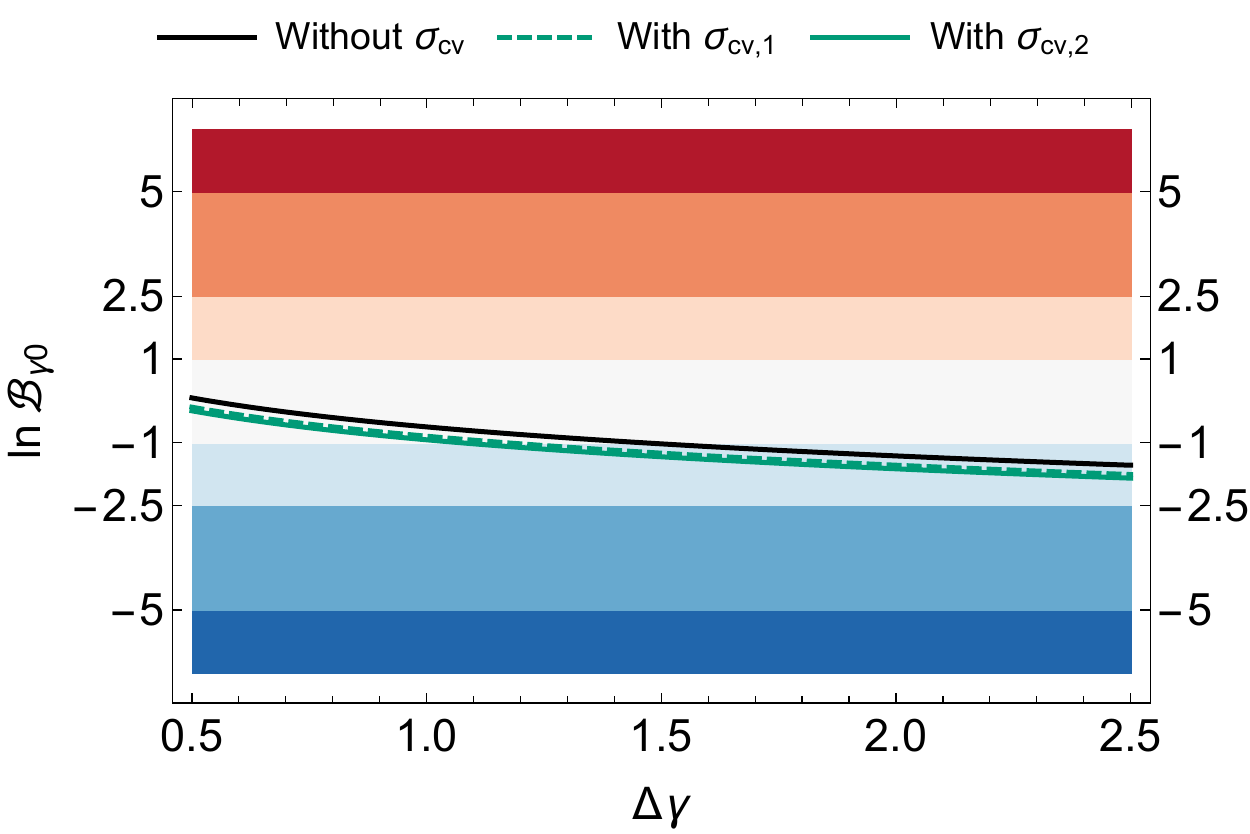}
\caption{Bayes factor as a function of the prior width. See section~\ref{result} for details.}
\label{fig:pg}
\end{figure}

\begin{figure}
\centering
\includegraphics[width=0.47\textwidth]{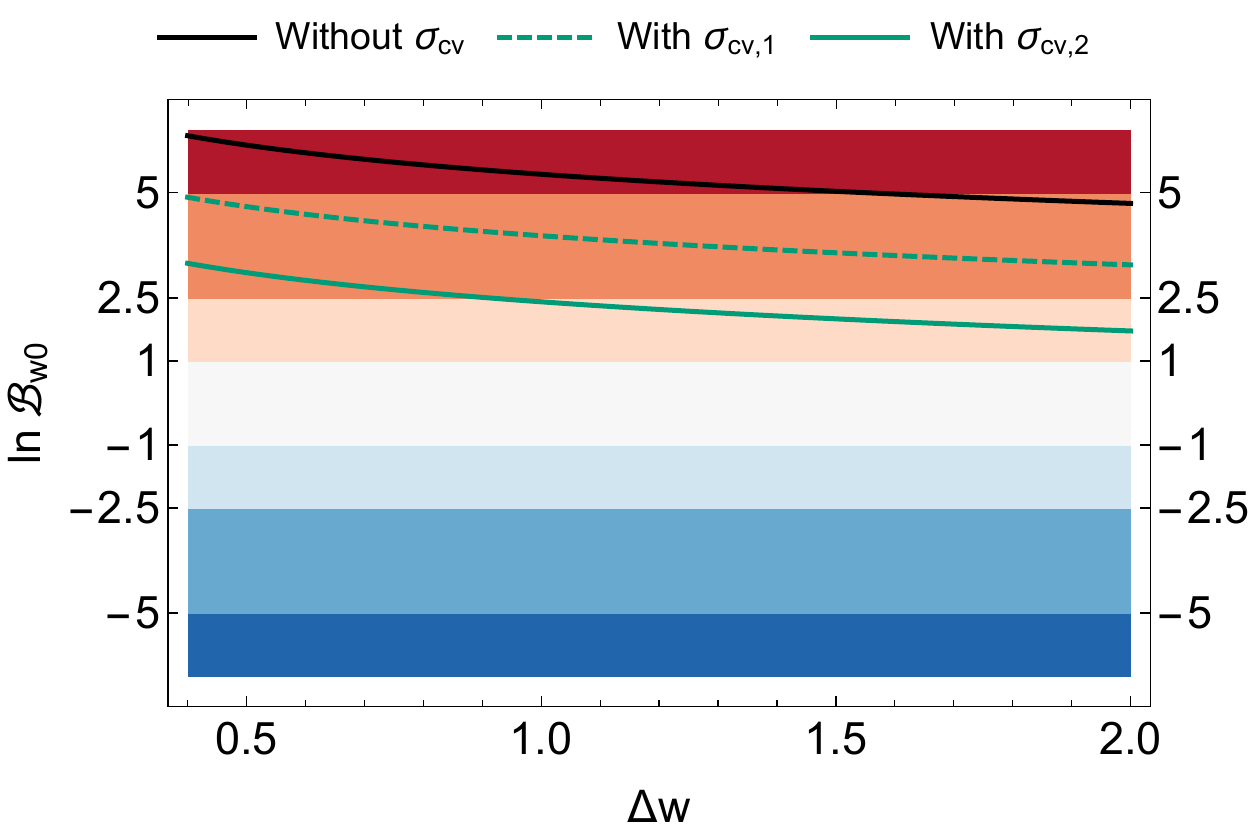}
\caption{Bayes factor as a function of the prior width. See section~\ref{result} for details.}
\label{fig:pw}
\end{figure}

\begin{figure*}
\centering
\includegraphics[width=0.325\textwidth]{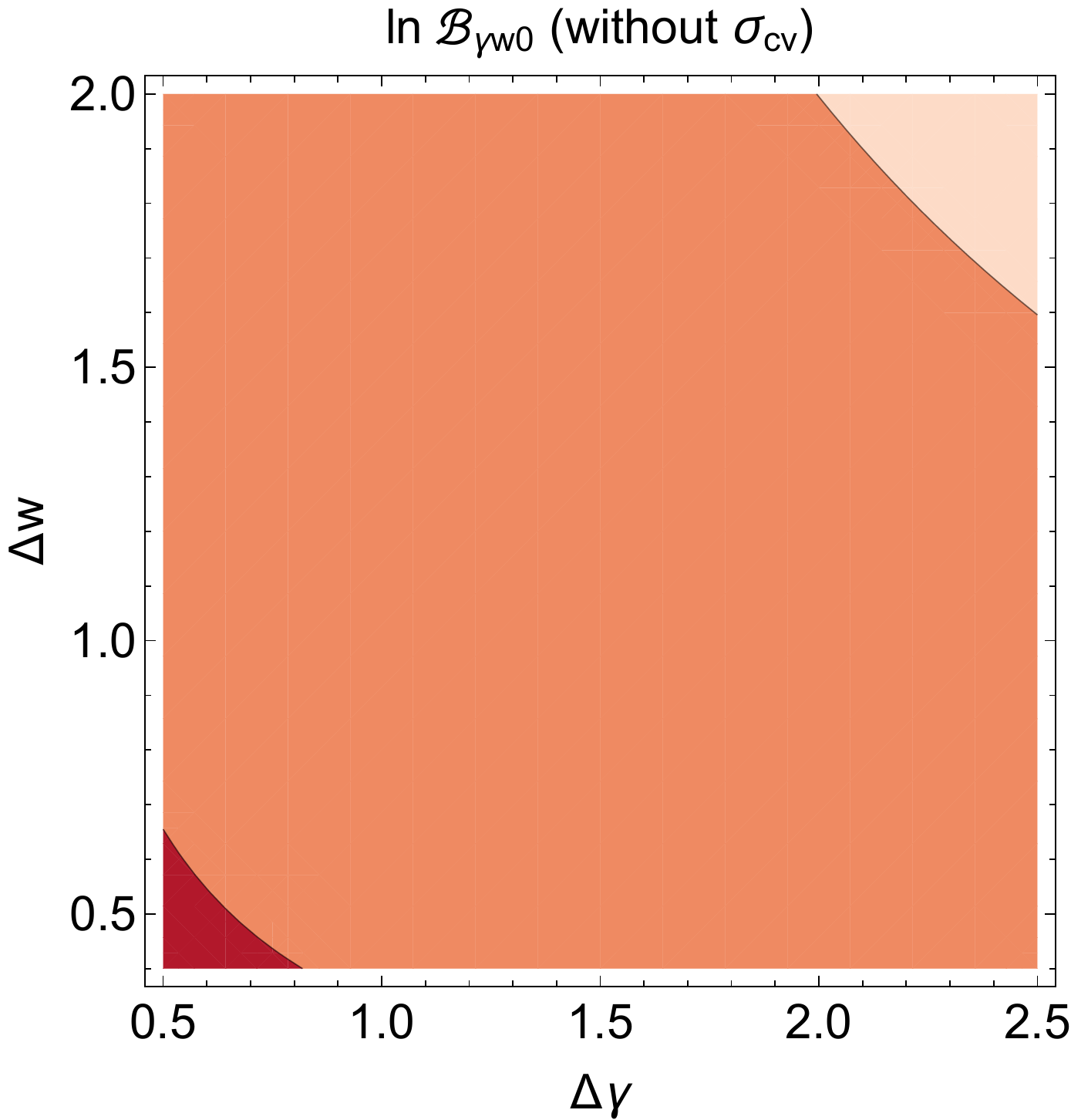}
\includegraphics[width=0.325\textwidth]{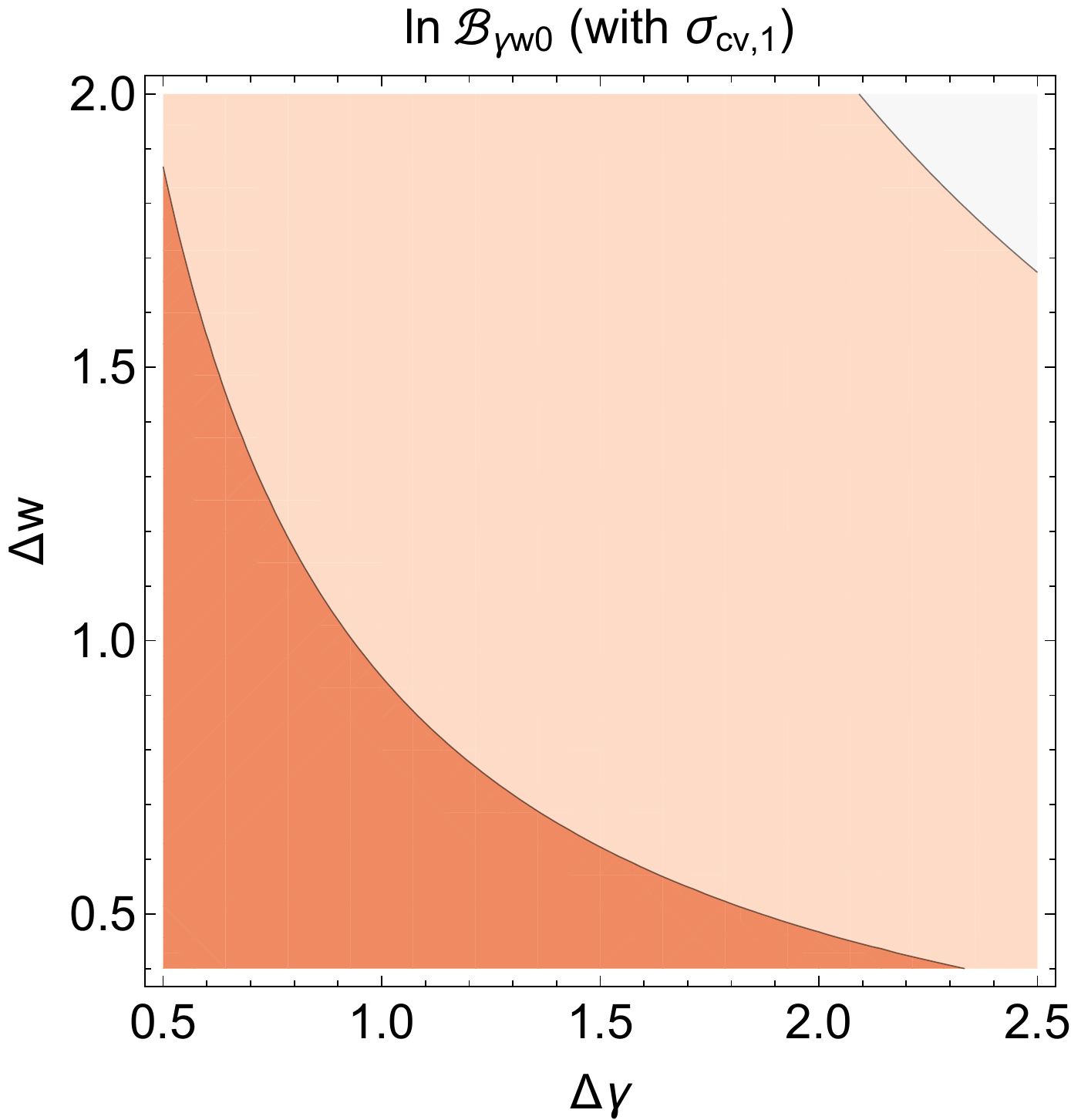}
\includegraphics[width=0.325\textwidth]{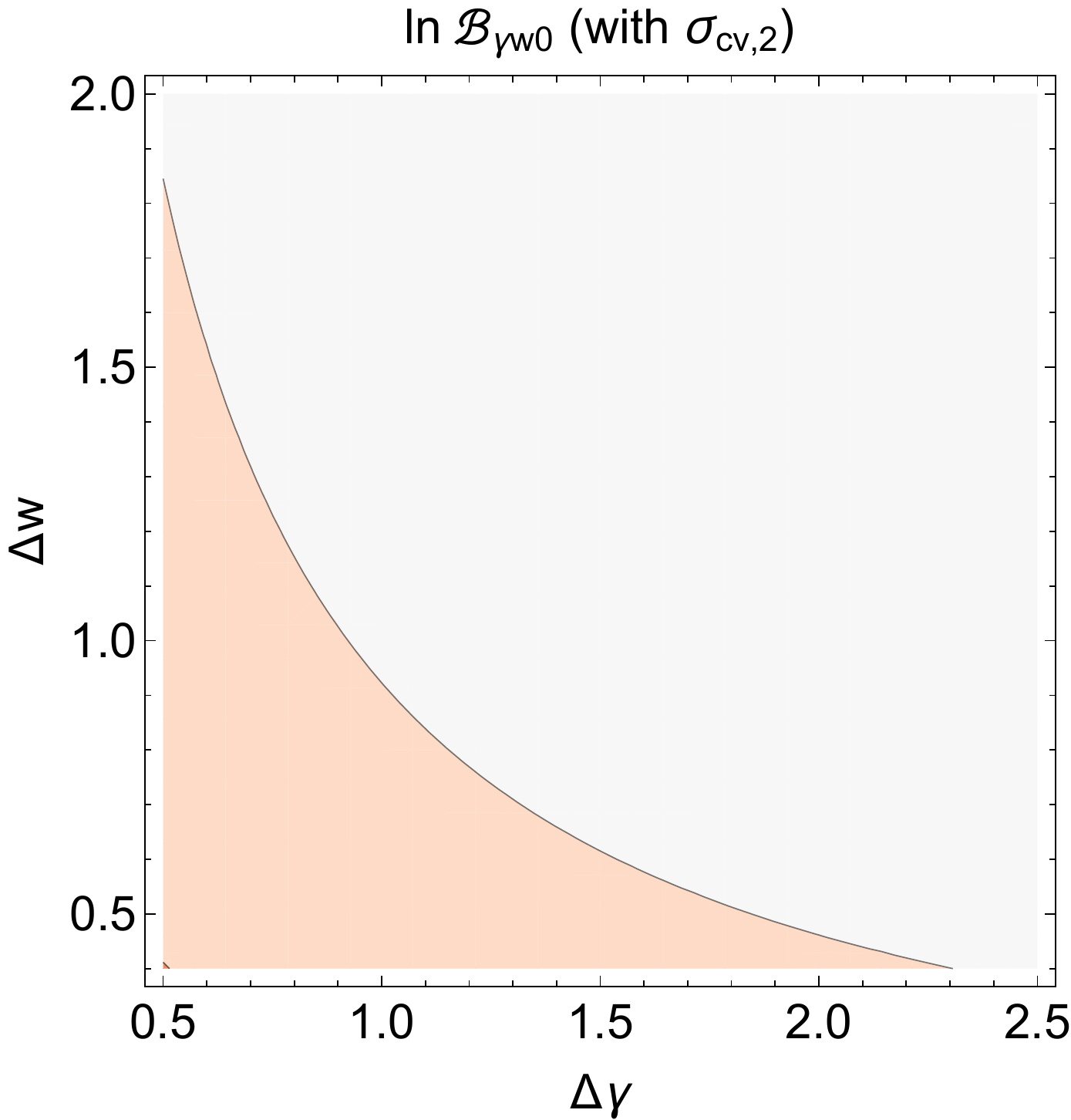}
\caption{Bayes factor as a function of the prior widths. See section~\ref{result} for details.}
\label{fig:pgw}
\end{figure*}

Similar behaviors follow the $\bigtriangleup$AIC and $\bigtriangleup$BIC differences (sixth and seventh columns of table \ref{tabRX}).
In particular, using the qualitative interpretations given in tables~\ref{tabAIC} and \ref{tabBIC} and neglecting $\sigma_{\rm cv}$ one concludes that $\Lambda$CDM is considerable less supported by data with respect to the $w$CDM model ($\bigtriangleup$AIC) and that there is a positive evidence against it ($\bigtriangleup$BIC).
However, if $\sigma_{\rm cv}$ is considered, the evidence in favor of $w$CDM model becomes a category weaker.
Therefore, the cosmic variance on $H_0$ not only shifts the constraints and improve the fit to the data but also changes model selection.
This is confirmed by figures~\ref{fig:pg}--\ref{fig:pgw} which show the Bayes factor, equation \eqref{eq:bayesfactor}, as a function of the prior widths. The colors are coded according to table~\ref{jeffreys}. The Bayes factor depends weakly (logarithmically) on the widths.
The widths ranges go from the minimum values necessary to close the unmarginalized posterior to 5 times the latter value.
Again, by including the cosmic variance in the analysis one goes from a strong evidence for models with the $w$ parameter to moderate evidence.
The impact on the $\gamma$ parameter is instead negligible.

\section{Conclusions}\label{conclu}

We have studied the impact of including the  cosmic variance on the Hubble constant on statistical inference--in particular in light of the $3.8\sigma$ tension on local $H_0$.
We considered the $\gamma$CDM, $w$CDM and $\gamma w$CDM parametric extensions of the standard model and the latest CMB, BAO, SNe Ia, RSD and $H_0$ data.

We showed that the systematic error from cosmic variance is, with little variation, approximately $\sigma_{\rm cv}\approx 0.88$ km s$^{-1}$ Mpc$^{-1}$ (1.2\% $H_0^{\text{loc}}$) when considering the redshift range $0.0233 \le z \le 0.15$ and $\sigma_{\rm cv}\approx 1.5$ km s$^{-1}$ Mpc$^{-1}$ (2.1\% $H_0^{\text{loc}}$) when considering the redshift range $0.01 \le z \le 0.15$.
The former range is used in the main part of the analysis by~\cite{Riess:2018byc} as it helps to reduce cosmic variance.
One may roughly estimate the error due to cosmic variance by assuming the latter values in equation \eqref{cv1} and \eqref{cv2}, without going through the method detailed in Section~\ref{gen}.

The inclusion of $\sigma_{\rm cv}$ lowers the tension and shifts the parameters correlated (directly or indirectly) with $H_0$. This produces important changes in the case of the $w$CDM model as the posterior is pushed towards non-phantom values.

Even more important are the implications regarding model selection.
We computed differences in $\chi^2_{\rm min}$, AIC and BIC, and the Bayes factor as a function of the prior widths, and we found that the alternative models with free equation of state $w$ lose their strong support when the cosmic variance is included. 
Indeed, models such as  $w$CDM  can accommodate a higher $H_0$ at the price of a phantom equation of state ($w<-1$).
This is the reason why the Bayes factor with respect to $\Lambda$CDM is so high (see Figure~\ref{fig:pw}).
Once the cosmic variance on $H_0$ is included in the analysis, there is less statistical gain in having a higher $H_0$ and the $w$CDM model is only moderately supported.
This can be interpreted as a volume effect, which is the quantitative formulation of the qualitative Occam's razor: as the uncertainty on $H_{0}$ increases it is more difficult to justify the parameter space volume associated to the extra parameter $w$.
While we analyzed only parametric extensions of the $\Lambda$CDM model, these conclusions (biased model selection) could  hold for more specific non-standard models that can accommodate a higher $H_0$.

As said earlier, the tension between global and local $H_0$ may favor non-standard models.
For this reason we think that it is safer to use a theoretical estimation of the cosmic variance which is not based on analyses carried out assuming the standard model (at least until the tension is well understood and explained).
While for the standard $\Lambda$CDM model it may be possible to constrain the local peculiar velocity flow with observations, this procedure is based on results (e.g., data analyses and simulations) that are not necessarily valid for non-standard models.
For example, correcting, as done in~\cite{Riess:2016jrr}, the individual SN redshifts for the local mass density as measured in flow maps may be correct within the $\Lambda$CDM model and it may correct potential biases on its parameters but it could bias model selection with respect to non-standard exotic models, which could feature a different growth of structures and a different cosmic variance.
According to our results, one should evaluate the cosmic variance on local $H_0$ for the models under consideration and include it in the error budget.
Neglecting its effect could potentially bias the conclusions of both parameter estimation and model selection.

Finally, it could be that cosmic variance has a minor role, that local determinations of $H_0$ already consider all possible sources of systematics and that CMB observations suffer from unaccounted-for systematics which bias the global $H_0$ towards lower values.
In order to exclude this possibility it will be crucial to determine $H_0$ at redshifts $0.2 \lesssim z \lesssim 0.5$~\cite{Tully:2016ppz,Feeney:2018mkj,daSilva:2018sim,Gomez-Valent:2018hwc}, that is, at scales at which cosmic variance is expected to be negligible.

\begin{acknowledgments}
It is a pleasure to thank Miguel Quartin, Oliver Piattella, Adam Riess and Wiliam Hipólito for useful comments and discussions.
DC thanks CAPES for  financial support.
VM thanks CNPq and FAPES for partial financial support. 
\end{acknowledgments}

\bibliography{biblio}

\begin{thebibliography}{74}%
\makeatletter
\providecommand \@ifxundefined [1]{%
 \@ifx{#1\undefined}
}%
\providecommand \@ifnum [1]{%
 \ifnum #1\expandafter \@firstoftwo
 \else \expandafter \@secondoftwo
 \fi
}%
\providecommand \@ifx [1]{%
 \ifx #1\expandafter \@firstoftwo
 \else \expandafter \@secondoftwo
 \fi
}%
\providecommand \natexlab [1]{#1}%
\providecommand \enquote  [1]{``#1''}%
\providecommand \bibnamefont  [1]{#1}%
\providecommand \bibfnamefont [1]{#1}%
\providecommand \citenamefont [1]{#1}%
\providecommand \href@noop [0]{\@secondoftwo}%
\providecommand \href [0]{\begingroup \@sanitize@url \@href}%
\providecommand \@href[1]{\@@startlink{#1}\@@href}%
\providecommand \@@href[1]{\endgroup#1\@@endlink}%
\providecommand \@sanitize@url [0]{\catcode `\\12\catcode `\$12\catcode
  `\&12\catcode `\#12\catcode `\^12\catcode `\_12\catcode `\%12\relax}%
\providecommand \@@startlink[1]{}%
\providecommand \@@endlink[0]{}%
\providecommand \url  [0]{\begingroup\@sanitize@url \@url }%
\providecommand \@url [1]{\endgroup\@href {#1}{\urlprefix }}%
\providecommand \urlprefix  [0]{URL }%
\providecommand \Eprint [0]{\href }%
\providecommand \doibase [0]{http://dx.doi.org/}%
\providecommand \selectlanguage [0]{\@gobble}%
\providecommand \bibinfo  [0]{\@secondoftwo}%
\providecommand \bibfield  [0]{\@secondoftwo}%
\providecommand \translation [1]{[#1]}%
\providecommand \BibitemOpen [0]{}%
\providecommand \bibitemStop [0]{}%
\providecommand \bibitemNoStop [0]{.\EOS\space}%
\providecommand \EOS [0]{\spacefactor3000\relax}%
\providecommand \BibitemShut  [1]{\csname bibitem#1\endcsname}%
\let\auto@bib@innerbib\@empty
\bibitem [{\citenamefont {Riess}\ \emph {et~al.}(2018)\citenamefont {Riess}
  \emph {et~al.}}]{Riess:2018byc}%
  \BibitemOpen
  \bibfield  {author} {\bibinfo {author} {\bibfnamefont {A.~G.}\ \bibnamefont
  {Riess}} \emph {et~al.},\ }\href@noop {} {\  (\bibinfo {year} {2018})},\
  \Eprint {http://arxiv.org/abs/1804.10655} {arXiv:1804.10655 [astro-ph.CO]}
  \BibitemShut {NoStop}%
\bibitem [{\citenamefont {Aghanim}\ \emph {et~al.}(2016)\citenamefont {Aghanim}
  \emph {et~al.}}]{Aghanim:2016yuo}%
  \BibitemOpen
  \bibfield  {author} {\bibinfo {author} {\bibfnamefont {N.}~\bibnamefont
  {Aghanim}} \emph {et~al.} (\bibinfo {collaboration} {Planck}),\ }\href
  {\doibase 10.1051/0004-6361/201628890} {\bibfield  {journal} {\bibinfo
  {journal} {Astron. Astrophys.}\ }\textbf {\bibinfo {volume} {596}},\ \bibinfo
  {pages} {A107} (\bibinfo {year} {2016})},\ \Eprint
  {http://arxiv.org/abs/1605.02985} {arXiv:1605.02985 [astro-ph.CO]}
  \BibitemShut {NoStop}%
\bibitem [{\citenamefont {Efstathiou}(2014)}]{Efstathiou:2013via}%
  \BibitemOpen
  \bibfield  {author} {\bibinfo {author} {\bibfnamefont {G.}~\bibnamefont
  {Efstathiou}},\ }\href {\doibase 10.1093/mnras/stu278} {\bibfield  {journal}
  {\bibinfo  {journal} {Mon. Not. Roy. Astron. Soc.}\ }\textbf {\bibinfo
  {volume} {440}},\ \bibinfo {pages} {1138} (\bibinfo {year} {2014})},\ \Eprint
  {http://arxiv.org/abs/1311.3461} {arXiv:1311.3461 [astro-ph.CO]} \BibitemShut
  {NoStop}%
\bibitem [{\citenamefont {Cardona}\ \emph {et~al.}(2017)\citenamefont
  {Cardona}, \citenamefont {Kunz},\ and\ \citenamefont
  {Pettorino}}]{Cardona:2016ems}%
  \BibitemOpen
  \bibfield  {author} {\bibinfo {author} {\bibfnamefont {W.}~\bibnamefont
  {Cardona}}, \bibinfo {author} {\bibfnamefont {M.}~\bibnamefont {Kunz}}, \
  and\ \bibinfo {author} {\bibfnamefont {V.}~\bibnamefont {Pettorino}},\ }\href
  {\doibase 10.1088/1475-7516/2017/03/056} {\bibfield  {journal} {\bibinfo
  {journal} {JCAP}\ }\textbf {\bibinfo {volume} {1703}},\ \bibinfo {pages}
  {056} (\bibinfo {year} {2017})},\ \Eprint {http://arxiv.org/abs/1611.06088}
  {arXiv:1611.06088 [astro-ph.CO]} \BibitemShut {NoStop}%
\bibitem [{\citenamefont {Zhang}\ \emph {et~al.}(2017)\citenamefont {Zhang},
  \citenamefont {Childress}, \citenamefont {Davis}, \citenamefont {Karpenka},
  \citenamefont {Lidman}, \citenamefont {Schmidt},\ and\ \citenamefont
  {Smith}}]{Zhang:2017aqn}%
  \BibitemOpen
  \bibfield  {author} {\bibinfo {author} {\bibfnamefont {B.~R.}\ \bibnamefont
  {Zhang}}, \bibinfo {author} {\bibfnamefont {M.~J.}\ \bibnamefont
  {Childress}}, \bibinfo {author} {\bibfnamefont {T.~M.}\ \bibnamefont
  {Davis}}, \bibinfo {author} {\bibfnamefont {N.~V.}\ \bibnamefont {Karpenka}},
  \bibinfo {author} {\bibfnamefont {C.}~\bibnamefont {Lidman}}, \bibinfo
  {author} {\bibfnamefont {B.~P.}\ \bibnamefont {Schmidt}}, \ and\ \bibinfo
  {author} {\bibfnamefont {M.}~\bibnamefont {Smith}},\ }\href {\doibase
  10.1093/mnras/stx1600} {\bibfield  {journal} {\bibinfo  {journal} {Mon. Not.
  Roy. Astron. Soc.}\ }\textbf {\bibinfo {volume} {471}},\ \bibinfo {pages}
  {2254} (\bibinfo {year} {2017})},\ \Eprint {http://arxiv.org/abs/1706.07573}
  {arXiv:1706.07573 [astro-ph.CO]} \BibitemShut {NoStop}%
\bibitem [{\citenamefont {Feeney}\ \emph {et~al.}(2017)\citenamefont {Feeney},
  \citenamefont {Mortlock},\ and\ \citenamefont {Dalmasso}}]{Feeney:2017sgx}%
  \BibitemOpen
  \bibfield  {author} {\bibinfo {author} {\bibfnamefont {S.~M.}\ \bibnamefont
  {Feeney}}, \bibinfo {author} {\bibfnamefont {D.~J.}\ \bibnamefont
  {Mortlock}}, \ and\ \bibinfo {author} {\bibfnamefont {N.}~\bibnamefont
  {Dalmasso}},\ }\href {\doibase 10.1093/mnras/sty418} {\  (\bibinfo {year}
  {2017}),\ 10.1093/mnras/sty418},\ \Eprint {http://arxiv.org/abs/1707.00007}
  {arXiv:1707.00007 [astro-ph.CO]} \BibitemShut {NoStop}%
\bibitem [{\citenamefont {Dhawan}\ \emph {et~al.}(2018)\citenamefont {Dhawan},
  \citenamefont {Jha},\ and\ \citenamefont {Leibundgut}}]{Dhawan:2017ywl}%
  \BibitemOpen
  \bibfield  {author} {\bibinfo {author} {\bibfnamefont {S.}~\bibnamefont
  {Dhawan}}, \bibinfo {author} {\bibfnamefont {S.~W.}\ \bibnamefont {Jha}}, \
  and\ \bibinfo {author} {\bibfnamefont {B.}~\bibnamefont {Leibundgut}},\
  }\href {\doibase 10.1051/0004-6361/201731501} {\bibfield  {journal} {\bibinfo
   {journal} {Astron. Astrophys.}\ }\textbf {\bibinfo {volume} {609}},\
  \bibinfo {pages} {A72} (\bibinfo {year} {2018})},\ \Eprint
  {http://arxiv.org/abs/1707.00715} {arXiv:1707.00715 [astro-ph.CO]}
  \BibitemShut {NoStop}%
\bibitem [{\citenamefont {Bernal}\ and\ \citenamefont
  {Peacock}(2018)}]{Bernal:2018cxc}%
  \BibitemOpen
  \bibfield  {author} {\bibinfo {author} {\bibfnamefont {J.~L.}\ \bibnamefont
  {Bernal}}\ and\ \bibinfo {author} {\bibfnamefont {J.~A.}\ \bibnamefont
  {Peacock}},\ }\href@noop {} {\  (\bibinfo {year} {2018})},\ \Eprint
  {http://arxiv.org/abs/1803.04470} {arXiv:1803.04470 [astro-ph.CO]}
  \BibitemShut {NoStop}%
\bibitem [{\citenamefont {Di~Valentino}\ \emph {et~al.}(2017)\citenamefont
  {Di~Valentino}, \citenamefont {Melchiorri},\ and\ \citenamefont
  {Mena}}]{DiValentino:2017iww}%
  \BibitemOpen
  \bibfield  {author} {\bibinfo {author} {\bibfnamefont {E.}~\bibnamefont
  {Di~Valentino}}, \bibinfo {author} {\bibfnamefont {A.}~\bibnamefont
  {Melchiorri}}, \ and\ \bibinfo {author} {\bibfnamefont {O.}~\bibnamefont
  {Mena}},\ }\href {\doibase 10.1103/PhysRevD.96.043503} {\bibfield  {journal}
  {\bibinfo  {journal} {Phys. Rev.}\ }\textbf {\bibinfo {volume} {D96}},\
  \bibinfo {pages} {043503} (\bibinfo {year} {2017})},\ \Eprint
  {http://arxiv.org/abs/1704.08342} {arXiv:1704.08342 [astro-ph.CO]}
  \BibitemShut {NoStop}%
\bibitem [{\citenamefont {Odderskov}\ \emph
  {et~al.}(2016{\natexlab{a}})\citenamefont {Odderskov}, \citenamefont
  {Baldi},\ and\ \citenamefont {Amendola}}]{Odderskov:2015fba}%
  \BibitemOpen
  \bibfield  {author} {\bibinfo {author} {\bibfnamefont {I.}~\bibnamefont
  {Odderskov}}, \bibinfo {author} {\bibfnamefont {M.}~\bibnamefont {Baldi}}, \
  and\ \bibinfo {author} {\bibfnamefont {L.}~\bibnamefont {Amendola}},\ }\href
  {\doibase 10.1088/1475-7516/2016/05/035} {\bibfield  {journal} {\bibinfo
  {journal} {JCAP}\ }\textbf {\bibinfo {volume} {1605}},\ \bibinfo {pages}
  {035} (\bibinfo {year} {2016}{\natexlab{a}})},\ \Eprint
  {http://arxiv.org/abs/1510.04314} {arXiv:1510.04314 [astro-ph.CO]}
  \BibitemShut {NoStop}%
\bibitem [{\citenamefont {Di~Valentino}\ \emph {et~al.}(2016)\citenamefont
  {Di~Valentino}, \citenamefont {Melchiorri},\ and\ \citenamefont
  {Silk}}]{DiValentino:2016hlg}%
  \BibitemOpen
  \bibfield  {author} {\bibinfo {author} {\bibfnamefont {E.}~\bibnamefont
  {Di~Valentino}}, \bibinfo {author} {\bibfnamefont {A.}~\bibnamefont
  {Melchiorri}}, \ and\ \bibinfo {author} {\bibfnamefont {J.}~\bibnamefont
  {Silk}},\ }\href {\doibase 10.1016/j.physletb.2016.08.043} {\bibfield
  {journal} {\bibinfo  {journal} {Phys. Lett.}\ }\textbf {\bibinfo {volume}
  {B761}},\ \bibinfo {pages} {242} (\bibinfo {year} {2016})},\ \Eprint
  {http://arxiv.org/abs/1606.00634} {arXiv:1606.00634 [astro-ph.CO]}
  \BibitemShut {NoStop}%
\bibitem [{\citenamefont {Di~Valentino}\ \emph {et~al.}(2018)\citenamefont
  {Di~Valentino}, \citenamefont {Linder},\ and\ \citenamefont
  {Melchiorri}}]{DiValentino:2017rcr}%
  \BibitemOpen
  \bibfield  {author} {\bibinfo {author} {\bibfnamefont {E.}~\bibnamefont
  {Di~Valentino}}, \bibinfo {author} {\bibfnamefont {E.~V.}\ \bibnamefont
  {Linder}}, \ and\ \bibinfo {author} {\bibfnamefont {A.}~\bibnamefont
  {Melchiorri}},\ }\href {\doibase 10.1103/PhysRevD.97.043528} {\bibfield
  {journal} {\bibinfo  {journal} {Phys. Rev.}\ }\textbf {\bibinfo {volume}
  {D97}},\ \bibinfo {pages} {043528} (\bibinfo {year} {2018})},\ \Eprint
  {http://arxiv.org/abs/1710.02153} {arXiv:1710.02153 [astro-ph.CO]}
  \BibitemShut {NoStop}%
\bibitem [{\citenamefont {Zhao}\ \emph {et~al.}(2017)\citenamefont {Zhao},
  \citenamefont {He}, \citenamefont {Zhang},\ and\ \citenamefont
  {Zhang}}]{Zhao:2017urm}%
  \BibitemOpen
  \bibfield  {author} {\bibinfo {author} {\bibfnamefont {M.-M.}\ \bibnamefont
  {Zhao}}, \bibinfo {author} {\bibfnamefont {D.-Z.}\ \bibnamefont {He}},
  \bibinfo {author} {\bibfnamefont {J.-F.}\ \bibnamefont {Zhang}}, \ and\
  \bibinfo {author} {\bibfnamefont {X.}~\bibnamefont {Zhang}},\ }\href
  {\doibase 10.1103/PhysRevD.96.043520} {\bibfield  {journal} {\bibinfo
  {journal} {Phys. Rev.}\ }\textbf {\bibinfo {volume} {D96}},\ \bibinfo {pages}
  {043520} (\bibinfo {year} {2017})},\ \Eprint
  {http://arxiv.org/abs/1703.08456} {arXiv:1703.08456 [astro-ph.CO]}
  \BibitemShut {NoStop}%
\bibitem [{\citenamefont {van Putten}(2017)}]{vanPutten:2017qte}%
  \BibitemOpen
  \bibfield  {author} {\bibinfo {author} {\bibfnamefont {M.~H. P.~M.}\
  \bibnamefont {van Putten}}\ }(\bibinfo {year} {2017})\ \Eprint
  {http://arxiv.org/abs/1707.02588} {arXiv:1707.02588 [astro-ph.CO]}
  \BibitemShut {NoStop}%
\bibitem [{\citenamefont {Solà}\ \emph {et~al.}(2017)\citenamefont {Solà},
  \citenamefont {Gómez-Valent},\ and\ \citenamefont
  {de~Cruz~Pérez}}]{Sola:2017znb}%
  \BibitemOpen
  \bibfield  {author} {\bibinfo {author} {\bibfnamefont {J.}~\bibnamefont
  {Solà}}, \bibinfo {author} {\bibfnamefont {A.}~\bibnamefont
  {Gómez-Valent}}, \ and\ \bibinfo {author} {\bibfnamefont {J.}~\bibnamefont
  {de~Cruz~Pérez}},\ }\href {\doibase 10.1016/j.physletb.2017.09.073}
  {\bibfield  {journal} {\bibinfo  {journal} {Phys. Lett.}\ }\textbf {\bibinfo
  {volume} {B774}},\ \bibinfo {pages} {317} (\bibinfo {year} {2017})},\ \Eprint
  {http://arxiv.org/abs/1705.06723} {arXiv:1705.06723 [astro-ph.CO]}
  \BibitemShut {NoStop}%
\bibitem [{\citenamefont {Huang}\ and\ \citenamefont
  {Wang}(2016)}]{Qing-Guo:2016ykt}%
  \BibitemOpen
  \bibfield  {author} {\bibinfo {author} {\bibfnamefont {Q.-G.}\ \bibnamefont
  {Huang}}\ and\ \bibinfo {author} {\bibfnamefont {K.}~\bibnamefont {Wang}},\
  }\href {\doibase 10.1140/epjc/s10052-016-4352-x} {\bibfield  {journal}
  {\bibinfo  {journal} {Eur. Phys. J.}\ }\textbf {\bibinfo {volume} {C76}},\
  \bibinfo {pages} {506} (\bibinfo {year} {2016})},\ \Eprint
  {http://arxiv.org/abs/1606.05965} {arXiv:1606.05965 [astro-ph.CO]}
  \BibitemShut {NoStop}%
\bibitem [{\citenamefont {{Turner}}\ \emph
  {et~al.}(1992{\natexlab{a}})\citenamefont {{Turner}}, \citenamefont {{Cen}},\
  and\ \citenamefont {{Ostriker}}}]{1992AJ....103.1427T}%
  \BibitemOpen
  \bibfield  {author} {\bibinfo {author} {\bibfnamefont {E.~L.}\ \bibnamefont
  {{Turner}}}, \bibinfo {author} {\bibfnamefont {R.}~\bibnamefont {{Cen}}}, \
  and\ \bibinfo {author} {\bibfnamefont {J.~P.}\ \bibnamefont {{Ostriker}}},\
  }\href {\doibase 10.1086/116156} {\bibfield  {journal} {\bibinfo  {journal}
  {Astron.J.}\ }\textbf {\bibinfo {volume} {103}},\ \bibinfo {pages} {1427}
  (\bibinfo {year} {1992}{\natexlab{a}})}\BibitemShut {NoStop}%
\bibitem [{\citenamefont {Suto}\ \emph {et~al.}(1995)\citenamefont {Suto},
  \citenamefont {Suginohara},\ and\ \citenamefont {Inagaki}}]{Suto:1994kd}%
  \BibitemOpen
  \bibfield  {author} {\bibinfo {author} {\bibfnamefont {Y.}~\bibnamefont
  {Suto}}, \bibinfo {author} {\bibfnamefont {T.}~\bibnamefont {Suginohara}}, \
  and\ \bibinfo {author} {\bibfnamefont {Y.}~\bibnamefont {Inagaki}},\ }\href
  {\doibase 10.1143/PTP.93.839} {\bibfield  {journal} {\bibinfo  {journal}
  {Prog.Theor.Phys.}\ }\textbf {\bibinfo {volume} {93}},\ \bibinfo {pages}
  {839} (\bibinfo {year} {1995})},\ \Eprint
  {http://arxiv.org/abs/astro-ph/9412090} {arXiv:astro-ph/9412090 [astro-ph]}
  \BibitemShut {NoStop}%
\bibitem [{\citenamefont {Shi}\ \emph {et~al.}(1996)\citenamefont {Shi},
  \citenamefont {Widrow},\ and\ \citenamefont {Dursi}}]{Shi:1995nq}%
  \BibitemOpen
  \bibfield  {author} {\bibinfo {author} {\bibfnamefont {X.}~\bibnamefont
  {Shi}}, \bibinfo {author} {\bibfnamefont {L.~M.}\ \bibnamefont {Widrow}}, \
  and\ \bibinfo {author} {\bibfnamefont {L.~J.}\ \bibnamefont {Dursi}},\
  }\href@noop {} {\bibfield  {journal} {\bibinfo  {journal}
  {Mon.Not.Roy.Astron.Soc.}\ }\textbf {\bibinfo {volume} {281}},\ \bibinfo
  {pages} {565} (\bibinfo {year} {1996})},\ \Eprint
  {http://arxiv.org/abs/astro-ph/9506120} {arXiv:astro-ph/9506120 [astro-ph]}
  \BibitemShut {NoStop}%
\bibitem [{\citenamefont {Shi}\ and\ \citenamefont
  {Turner}(1998)}]{Shi:1997aa}%
  \BibitemOpen
  \bibfield  {author} {\bibinfo {author} {\bibfnamefont {X.-D.}\ \bibnamefont
  {Shi}}\ and\ \bibinfo {author} {\bibfnamefont {M.~S.}\ \bibnamefont
  {Turner}},\ }\href {\doibase 10.1086/305169} {\bibfield  {journal} {\bibinfo
  {journal} {Astrophys.J.}\ }\textbf {\bibinfo {volume} {493}},\ \bibinfo
  {pages} {519} (\bibinfo {year} {1998})},\ \Eprint
  {http://arxiv.org/abs/astro-ph/9707101} {arXiv:astro-ph/9707101 [astro-ph]}
  \BibitemShut {NoStop}%
\bibitem [{\citenamefont {Wang}\ \emph {et~al.}(1998)\citenamefont {Wang},
  \citenamefont {Spergel},\ and\ \citenamefont {Turner}}]{Wang:1997tp}%
  \BibitemOpen
  \bibfield  {author} {\bibinfo {author} {\bibfnamefont {Y.}~\bibnamefont
  {Wang}}, \bibinfo {author} {\bibfnamefont {D.~N.}\ \bibnamefont {Spergel}}, \
  and\ \bibinfo {author} {\bibfnamefont {E.~L.}\ \bibnamefont {Turner}},\
  }\href {\doibase 10.1086/305539} {\bibfield  {journal} {\bibinfo  {journal}
  {Astrophys.J.}\ }\textbf {\bibinfo {volume} {498}},\ \bibinfo {pages} {1}
  (\bibinfo {year} {1998})},\ \Eprint {http://arxiv.org/abs/astro-ph/9708014}
  {arXiv:astro-ph/9708014 [astro-ph]} \BibitemShut {NoStop}%
\bibitem [{\citenamefont {Zehavi}\ \emph {et~al.}(1998)\citenamefont {Zehavi},
  \citenamefont {Riess}, \citenamefont {Kirshner},\ and\ \citenamefont
  {Dekel}}]{Zehavi:1998gz}%
  \BibitemOpen
  \bibfield  {author} {\bibinfo {author} {\bibfnamefont {I.}~\bibnamefont
  {Zehavi}}, \bibinfo {author} {\bibfnamefont {A.~G.}\ \bibnamefont {Riess}},
  \bibinfo {author} {\bibfnamefont {R.~P.}\ \bibnamefont {Kirshner}}, \ and\
  \bibinfo {author} {\bibfnamefont {A.}~\bibnamefont {Dekel}},\ }\href
  {\doibase 10.1086/306015} {\bibfield  {journal} {\bibinfo  {journal}
  {Astrophys.J.}\ }\textbf {\bibinfo {volume} {503}},\ \bibinfo {pages} {483}
  (\bibinfo {year} {1998})},\ \Eprint {http://arxiv.org/abs/astro-ph/9802252}
  {arXiv:astro-ph/9802252 [astro-ph]} \BibitemShut {NoStop}%
\bibitem [{\citenamefont {Giovanelli}\ \emph {et~al.}(1999)\citenamefont
  {Giovanelli}, \citenamefont {Dale}, \citenamefont {Haynes}, \citenamefont
  {Hardy},\ and\ \citenamefont {Campusano}}]{Giovanelli:1999xp}%
  \BibitemOpen
  \bibfield  {author} {\bibinfo {author} {\bibfnamefont {R.}~\bibnamefont
  {Giovanelli}}, \bibinfo {author} {\bibfnamefont {D.}~\bibnamefont {Dale}},
  \bibinfo {author} {\bibfnamefont {M.}~\bibnamefont {Haynes}}, \bibinfo
  {author} {\bibfnamefont {E.}~\bibnamefont {Hardy}}, \ and\ \bibinfo {author}
  {\bibfnamefont {L.}~\bibnamefont {Campusano}},\ }\href {\doibase
  10.1086/307906} {\bibfield  {journal} {\bibinfo  {journal} {Astrophys.J.}\
  }\textbf {\bibinfo {volume} {525}},\ \bibinfo {pages} {25} (\bibinfo {year}
  {1999})},\ \Eprint {http://arxiv.org/abs/astro-ph/9906362}
  {arXiv:astro-ph/9906362 [astro-ph]} \BibitemShut {NoStop}%
\bibitem [{\citenamefont {Marra}\ \emph {et~al.}(2013)\citenamefont {Marra},
  \citenamefont {Amendola}, \citenamefont {Sawicki},\ and\ \citenamefont
  {Valkenburg}}]{Marra:2013rba}%
  \BibitemOpen
  \bibfield  {author} {\bibinfo {author} {\bibfnamefont {V.}~\bibnamefont
  {Marra}}, \bibinfo {author} {\bibfnamefont {L.}~\bibnamefont {Amendola}},
  \bibinfo {author} {\bibfnamefont {I.}~\bibnamefont {Sawicki}}, \ and\
  \bibinfo {author} {\bibfnamefont {W.}~\bibnamefont {Valkenburg}},\ }\href
  {\doibase 10.1103/PhysRevLett.110.241305} {\bibfield  {journal} {\bibinfo
  {journal} {Phys. Rev. Lett.}\ }\textbf {\bibinfo {volume} {110}},\ \bibinfo
  {pages} {241305} (\bibinfo {year} {2013})},\ \Eprint
  {http://arxiv.org/abs/1303.3121} {arXiv:1303.3121 [astro-ph.CO]} \BibitemShut
  {NoStop}%
\bibitem [{\citenamefont {Ben-Dayan}\ \emph {et~al.}(2014)\citenamefont
  {Ben-Dayan}, \citenamefont {Durrer}, \citenamefont {Marozzi},\ and\
  \citenamefont {Schwarz}}]{Ben-Dayan:2014swa}%
  \BibitemOpen
  \bibfield  {author} {\bibinfo {author} {\bibfnamefont {I.}~\bibnamefont
  {Ben-Dayan}}, \bibinfo {author} {\bibfnamefont {R.}~\bibnamefont {Durrer}},
  \bibinfo {author} {\bibfnamefont {G.}~\bibnamefont {Marozzi}}, \ and\
  \bibinfo {author} {\bibfnamefont {D.~J.}\ \bibnamefont {Schwarz}},\ }\href
  {\doibase 10.1103/PhysRevLett.112.221301} {\bibfield  {journal} {\bibinfo
  {journal} {Phys. Rev. Lett.}\ }\textbf {\bibinfo {volume} {112}},\ \bibinfo
  {pages} {221301} (\bibinfo {year} {2014})},\ \Eprint
  {http://arxiv.org/abs/1401.7973} {arXiv:1401.7973 [astro-ph.CO]} \BibitemShut
  {NoStop}%
\bibitem [{\citenamefont {Keenan}\ \emph {et~al.}(2013)\citenamefont {Keenan},
  \citenamefont {Barger},\ and\ \citenamefont {Cowie}}]{Keenan:2013mfa}%
  \BibitemOpen
  \bibfield  {author} {\bibinfo {author} {\bibfnamefont {R.~C.}\ \bibnamefont
  {Keenan}}, \bibinfo {author} {\bibfnamefont {A.~J.}\ \bibnamefont {Barger}},
  \ and\ \bibinfo {author} {\bibfnamefont {L.~L.}\ \bibnamefont {Cowie}},\
  }\href {\doibase 10.1088/0004-637X/775/1/62} {\bibfield  {journal} {\bibinfo
  {journal} {Astrophys. J.}\ }\textbf {\bibinfo {volume} {775}},\ \bibinfo
  {pages} {62} (\bibinfo {year} {2013})},\ \Eprint
  {http://arxiv.org/abs/1304.2884} {arXiv:1304.2884 [astro-ph.CO]} \BibitemShut
  {NoStop}%
\bibitem [{\citenamefont {Redlich}\ \emph {et~al.}(2014)\citenamefont
  {Redlich}, \citenamefont {Bolejko}, \citenamefont {Meyer}, \citenamefont
  {Lewis},\ and\ \citenamefont {Bartelmann}}]{Redlich:2014gga}%
  \BibitemOpen
  \bibfield  {author} {\bibinfo {author} {\bibfnamefont {M.}~\bibnamefont
  {Redlich}}, \bibinfo {author} {\bibfnamefont {K.}~\bibnamefont {Bolejko}},
  \bibinfo {author} {\bibfnamefont {S.}~\bibnamefont {Meyer}}, \bibinfo
  {author} {\bibfnamefont {G.~F.}\ \bibnamefont {Lewis}}, \ and\ \bibinfo
  {author} {\bibfnamefont {M.}~\bibnamefont {Bartelmann}},\ }\href {\doibase
  10.1051/0004-6361/201424553} {\bibfield  {journal} {\bibinfo  {journal}
  {Astron. Astrophys.}\ }\textbf {\bibinfo {volume} {570}},\ \bibinfo {pages}
  {A63} (\bibinfo {year} {2014})},\ \Eprint {http://arxiv.org/abs/1408.1872}
  {arXiv:1408.1872 [astro-ph.CO]} \BibitemShut {NoStop}%
\bibitem [{\citenamefont {Bengaly}(2016)}]{Bengaly:2015nwa}%
  \BibitemOpen
  \bibfield  {author} {\bibinfo {author} {\bibfnamefont {C.~A.~P.}\
  \bibnamefont {Bengaly}, \bibfnamefont {Jr.}},\ }\href {\doibase
  10.1088/1475-7516/2016/04/036} {\bibfield  {journal} {\bibinfo  {journal}
  {JCAP}\ }\textbf {\bibinfo {volume} {1604}},\ \bibinfo {pages} {036}
  (\bibinfo {year} {2016})},\ \Eprint {http://arxiv.org/abs/1510.05545}
  {arXiv:1510.05545 [astro-ph.CO]} \BibitemShut {NoStop}%
\bibitem [{\citenamefont {Hoscheit}\ and\ \citenamefont
  {Barger}(2018)}]{Hoscheit:2018nfl}%
  \BibitemOpen
  \bibfield  {author} {\bibinfo {author} {\bibfnamefont {B.~L.}\ \bibnamefont
  {Hoscheit}}\ and\ \bibinfo {author} {\bibfnamefont {A.~J.}\ \bibnamefont
  {Barger}},\ }\href {\doibase 10.3847/1538-4357/aaa59b} {\bibfield  {journal}
  {\bibinfo  {journal} {Astrophys. J.}\ }\textbf {\bibinfo {volume} {854}},\
  \bibinfo {pages} {46} (\bibinfo {year} {2018})},\ \Eprint
  {http://arxiv.org/abs/1801.01890} {arXiv:1801.01890 [astro-ph.CO]}
  \BibitemShut {NoStop}%
\bibitem [{\citenamefont {Wojtak}\ \emph {et~al.}(2014)\citenamefont {Wojtak},
  \citenamefont {Knebe}, \citenamefont {Watson}, \citenamefont {Iliev},
  \citenamefont {Heß}, \citenamefont {Rapetti}, \citenamefont {Yepes},\ and\
  \citenamefont {Gottlöber}}]{Wojtak:2013gda}%
  \BibitemOpen
  \bibfield  {author} {\bibinfo {author} {\bibfnamefont {R.}~\bibnamefont
  {Wojtak}}, \bibinfo {author} {\bibfnamefont {A.}~\bibnamefont {Knebe}},
  \bibinfo {author} {\bibfnamefont {W.~A.}\ \bibnamefont {Watson}}, \bibinfo
  {author} {\bibfnamefont {I.~T.}\ \bibnamefont {Iliev}}, \bibinfo {author}
  {\bibfnamefont {S.}~\bibnamefont {Heß}}, \bibinfo {author} {\bibfnamefont
  {D.}~\bibnamefont {Rapetti}}, \bibinfo {author} {\bibfnamefont
  {G.}~\bibnamefont {Yepes}}, \ and\ \bibinfo {author} {\bibfnamefont
  {S.}~\bibnamefont {Gottlöber}},\ }\href {\doibase 10.1093/mnras/stt2321}
  {\bibfield  {journal} {\bibinfo  {journal} {Mon. Not. Roy. Astron. Soc.}\
  }\textbf {\bibinfo {volume} {438}},\ \bibinfo {pages} {1805} (\bibinfo {year}
  {2014})},\ \Eprint {http://arxiv.org/abs/1312.0276} {arXiv:1312.0276
  [astro-ph.CO]} \BibitemShut {NoStop}%
\bibitem [{\citenamefont {Heß}\ and\ \citenamefont
  {Kitaura}(2016)}]{Hess:2014yka}%
  \BibitemOpen
  \bibfield  {author} {\bibinfo {author} {\bibfnamefont {S.}~\bibnamefont
  {Heß}}\ and\ \bibinfo {author} {\bibfnamefont {F.-S.}\ \bibnamefont
  {Kitaura}},\ }\href {\doibase 10.1093/mnras/stv2928} {\bibfield  {journal}
  {\bibinfo  {journal} {Mon. Not. Roy. Astron. Soc.}\ }\textbf {\bibinfo
  {volume} {456}},\ \bibinfo {pages} {4247} (\bibinfo {year} {2016})},\ \Eprint
  {http://arxiv.org/abs/1412.7310} {arXiv:1412.7310 [astro-ph.CO]} \BibitemShut
  {NoStop}%
\bibitem [{\citenamefont {Odderskov}\ \emph {et~al.}(2014)\citenamefont
  {Odderskov}, \citenamefont {Hannestad},\ and\ \citenamefont
  {Haugbølle}}]{Odderskov:2014hqa}%
  \BibitemOpen
  \bibfield  {author} {\bibinfo {author} {\bibfnamefont {I.}~\bibnamefont
  {Odderskov}}, \bibinfo {author} {\bibfnamefont {S.}~\bibnamefont
  {Hannestad}}, \ and\ \bibinfo {author} {\bibfnamefont {T.}~\bibnamefont
  {Haugbølle}},\ }\href {\doibase 10.1088/1475-7516/2014/10/028} {\bibfield
  {journal} {\bibinfo  {journal} {JCAP}\ }\textbf {\bibinfo {volume} {1410}},\
  \bibinfo {pages} {028} (\bibinfo {year} {2014})},\ \Eprint
  {http://arxiv.org/abs/1407.7364} {arXiv:1407.7364 [astro-ph.CO]} \BibitemShut
  {NoStop}%
\bibitem [{\citenamefont {Odderskov}\ \emph
  {et~al.}(2016{\natexlab{b}})\citenamefont {Odderskov}, \citenamefont
  {Koksbang},\ and\ \citenamefont {Hannestad}}]{Odderskov:2016rro}%
  \BibitemOpen
  \bibfield  {author} {\bibinfo {author} {\bibfnamefont {I.}~\bibnamefont
  {Odderskov}}, \bibinfo {author} {\bibfnamefont {S.~M.}\ \bibnamefont
  {Koksbang}}, \ and\ \bibinfo {author} {\bibfnamefont {S.}~\bibnamefont
  {Hannestad}},\ }\href {\doibase 10.1088/1475-7516/2016/02/001} {\bibfield
  {journal} {\bibinfo  {journal} {JCAP}\ }\textbf {\bibinfo {volume} {1602}},\
  \bibinfo {pages} {001} (\bibinfo {year} {2016}{\natexlab{b}})},\ \Eprint
  {http://arxiv.org/abs/1601.07356} {arXiv:1601.07356 [astro-ph.CO]}
  \BibitemShut {NoStop}%
\bibitem [{\citenamefont {Wu}\ and\ \citenamefont
  {Huterer}(2017)}]{Wu:2017fpr}%
  \BibitemOpen
  \bibfield  {author} {\bibinfo {author} {\bibfnamefont {H.-Y.}\ \bibnamefont
  {Wu}}\ and\ \bibinfo {author} {\bibfnamefont {D.}~\bibnamefont {Huterer}},\
  }\href {\doibase 10.1093/mnras/stx1967} {\bibfield  {journal} {\bibinfo
  {journal} {Mon. Not. Roy. Astron. Soc.}\ }\textbf {\bibinfo {volume} {471}},\
  \bibinfo {pages} {4946} (\bibinfo {year} {2017})},\ \Eprint
  {http://arxiv.org/abs/1706.09723} {arXiv:1706.09723 [astro-ph.CO]}
  \BibitemShut {NoStop}%
\bibitem [{\citenamefont {Odderskov}\ \emph {et~al.}(2017)\citenamefont
  {Odderskov}, \citenamefont {Hannestad},\ and\ \citenamefont
  {Brandbyge}}]{Odderskov:2017ivg}%
  \BibitemOpen
  \bibfield  {author} {\bibinfo {author} {\bibfnamefont {I.}~\bibnamefont
  {Odderskov}}, \bibinfo {author} {\bibfnamefont {S.}~\bibnamefont
  {Hannestad}}, \ and\ \bibinfo {author} {\bibfnamefont {J.}~\bibnamefont
  {Brandbyge}},\ }\href {\doibase 10.1088/1475-7516/2017/03/022} {\bibfield
  {journal} {\bibinfo  {journal} {JCAP}\ }\textbf {\bibinfo {volume} {1703}},\
  \bibinfo {pages} {022} (\bibinfo {year} {2017})},\ \Eprint
  {http://arxiv.org/abs/1701.05391} {arXiv:1701.05391 [astro-ph.CO]}
  \BibitemShut {NoStop}%
\bibitem [{\citenamefont {Kraljic}\ and\ \citenamefont
  {Sarkar}(2016)}]{Kraljic:2016acj}%
  \BibitemOpen
  \bibfield  {author} {\bibinfo {author} {\bibfnamefont {D.}~\bibnamefont
  {Kraljic}}\ and\ \bibinfo {author} {\bibfnamefont {S.}~\bibnamefont
  {Sarkar}},\ }\href {\doibase 10.1088/1475-7516/2016/10/016} {\bibfield
  {journal} {\bibinfo  {journal} {JCAP}\ }\textbf {\bibinfo {volume} {1610}},\
  \bibinfo {pages} {016} (\bibinfo {year} {2016})},\ \Eprint
  {http://arxiv.org/abs/1607.07377} {arXiv:1607.07377 [astro-ph.CO]}
  \BibitemShut {NoStop}%
\bibitem [{\citenamefont {Hellwing}\ \emph {et~al.}(2017)\citenamefont
  {Hellwing}, \citenamefont {Nusser}, \citenamefont {Feix},\ and\ \citenamefont
  {Bilicki}}]{Hellwing:2016pdl}%
  \BibitemOpen
  \bibfield  {author} {\bibinfo {author} {\bibfnamefont {W.~A.}\ \bibnamefont
  {Hellwing}}, \bibinfo {author} {\bibfnamefont {A.}~\bibnamefont {Nusser}},
  \bibinfo {author} {\bibfnamefont {M.}~\bibnamefont {Feix}}, \ and\ \bibinfo
  {author} {\bibfnamefont {M.}~\bibnamefont {Bilicki}},\ }\href {\doibase
  10.1093/mnras/stx213} {\bibfield  {journal} {\bibinfo  {journal} {Mon. Not.
  Roy. Astron. Soc.}\ }\textbf {\bibinfo {volume} {467}},\ \bibinfo {pages}
  {2787} (\bibinfo {year} {2017})},\ \Eprint {http://arxiv.org/abs/1609.07120}
  {arXiv:1609.07120 [astro-ph.CO]} \BibitemShut {NoStop}%
\bibitem [{\citenamefont {{Akaike}}(1974)}]{Akaike1974}%
  \BibitemOpen
  \bibfield  {author} {\bibinfo {author} {\bibfnamefont {H.}~\bibnamefont
  {{Akaike}}},\ }\href@noop {} {\bibfield  {journal} {\bibinfo  {journal} {IEEE
  Transactions on Automatic Control}\ }\textbf {\bibinfo {volume} {19}},\
  \bibinfo {pages} {716} (\bibinfo {year} {1974})}\BibitemShut {NoStop}%
\bibitem [{\citenamefont {Schwarz}(1978)}]{Schwarz:1978tpv}%
  \BibitemOpen
  \bibfield  {author} {\bibinfo {author} {\bibfnamefont {G.}~\bibnamefont
  {Schwarz}},\ }\href@noop {} {\bibfield  {journal} {\bibinfo  {journal}
  {Annals Statist.}\ }\textbf {\bibinfo {volume} {6}},\ \bibinfo {pages} {461}
  (\bibinfo {year} {1978})}\BibitemShut {NoStop}%
\bibitem [{\citenamefont {Joudaki}\ \emph {et~al.}(2017)\citenamefont {Joudaki}
  \emph {et~al.}}]{Joudaki:2016kym}%
  \BibitemOpen
  \bibfield  {author} {\bibinfo {author} {\bibfnamefont {S.}~\bibnamefont
  {Joudaki}} \emph {et~al.},\ }\href {\doibase 10.1093/mnras/stx998} {\bibfield
   {journal} {\bibinfo  {journal} {Mon. Not. Roy. Astron. Soc.}\ }\textbf
  {\bibinfo {volume} {471}},\ \bibinfo {pages} {1259} (\bibinfo {year}
  {2017})},\ \Eprint {http://arxiv.org/abs/1610.04606} {arXiv:1610.04606
  [astro-ph.CO]} \BibitemShut {NoStop}%
\bibitem [{\citenamefont {Lin}\ and\ \citenamefont
  {Ishak}(2017)}]{Lin:2017ikq}%
  \BibitemOpen
  \bibfield  {author} {\bibinfo {author} {\bibfnamefont {W.}~\bibnamefont
  {Lin}}\ and\ \bibinfo {author} {\bibfnamefont {M.}~\bibnamefont {Ishak}},\
  }\href {\doibase 10.1103/PhysRevD.96.023532} {\bibfield  {journal} {\bibinfo
  {journal} {Phys. Rev.}\ }\textbf {\bibinfo {volume} {D96}},\ \bibinfo {pages}
  {023532} (\bibinfo {year} {2017})},\ \Eprint
  {http://arxiv.org/abs/1705.05303} {arXiv:1705.05303 [astro-ph.CO]}
  \BibitemShut {NoStop}%
\bibitem [{\citenamefont {Ade}\ \emph {et~al.}(2016{\natexlab{a}})\citenamefont
  {Ade} \emph {et~al.}}]{Ade:2015XIII}%
  \BibitemOpen
  \bibfield  {author} {\bibinfo {author} {\bibfnamefont {P.~A.~R.}\
  \bibnamefont {Ade}} \emph {et~al.} (\bibinfo {collaboration} {Planck}),\
  }\href {\doibase 10.1051/0004-6361/201525830} {\bibfield  {journal} {\bibinfo
   {journal} {Astron. Astrophys.}\ }\textbf {\bibinfo {volume} {594}},\
  \bibinfo {pages} {A13} (\bibinfo {year} {2016}{\natexlab{a}})},\ \Eprint
  {http://arxiv.org/abs/1502.01589} {arXiv:1502.01589 [astro-ph.CO]}
  \BibitemShut {NoStop}%
\bibitem [{\citenamefont {{Turner}}\ \emph
  {et~al.}(1992{\natexlab{b}})\citenamefont {{Turner}}, \citenamefont {{Cen}},\
  and\ \citenamefont {{Ostriker}}}]{1992AJTurner}%
  \BibitemOpen
  \bibfield  {author} {\bibinfo {author} {\bibfnamefont {E.~L.}\ \bibnamefont
  {{Turner}}}, \bibinfo {author} {\bibfnamefont {R.}~\bibnamefont {{Cen}}}, \
  and\ \bibinfo {author} {\bibfnamefont {J.~P.}\ \bibnamefont {{Ostriker}}},\
  }\href {\doibase 10.1086/116156} {\bibfield  {journal} {\bibinfo  {journal}
  {Astrophys. J.}\ }\textbf {\bibinfo {volume} {103}},\ \bibinfo {pages} {1427}
  (\bibinfo {year} {1992}{\natexlab{b}})}\BibitemShut {NoStop}%
\bibitem [{\citenamefont {Deng}\ and\ \citenamefont
  {Wei}(2018)}]{Deng:2018jrp}%
  \BibitemOpen
  \bibfield  {author} {\bibinfo {author} {\bibfnamefont {H.-K.}\ \bibnamefont
  {Deng}}\ and\ \bibinfo {author} {\bibfnamefont {H.}~\bibnamefont {Wei}},\
  }\href@noop {} {\  (\bibinfo {year} {2018})},\ \Eprint
  {http://arxiv.org/abs/1806.02773} {arXiv:1806.02773 [astro-ph.CO]}
  \BibitemShut {NoStop}%
\bibitem [{\citenamefont {Scolnic}\ \emph {et~al.}(2017)\citenamefont {Scolnic}
  \emph {et~al.}}]{Scolnic:2017caz}%
  \BibitemOpen
  \bibfield  {author} {\bibinfo {author} {\bibfnamefont {D.~M.}\ \bibnamefont
  {Scolnic}} \emph {et~al.},\ }\href {\doibase 10.17909/T95Q4X} {\  (\bibinfo
  {year} {2017}),\ 10.17909/T95Q4X},\ \Eprint {http://arxiv.org/abs/1710.00845}
  {arXiv:1710.00845 [astro-ph.CO]} \BibitemShut {NoStop}%
\bibitem [{\citenamefont {Peebles}(1980)}]{peebles1980large}%
  \BibitemOpen
  \bibfield  {author} {\bibinfo {author} {\bibfnamefont {P.~J.~E.}\
  \bibnamefont {Peebles}},\ }\href@noop {} {\emph {\bibinfo {title} {The
  large-scale structure of the universe}}}\ (\bibinfo  {publisher} {Princeton
  university press},\ \bibinfo {year} {1980})\BibitemShut {NoStop}%
\bibitem [{\citenamefont {Wang}\ and\ \citenamefont
  {Steinhardt}(1998)}]{Wang:1998gt}%
  \BibitemOpen
  \bibfield  {author} {\bibinfo {author} {\bibfnamefont {L.-M.}\ \bibnamefont
  {Wang}}\ and\ \bibinfo {author} {\bibfnamefont {P.~J.}\ \bibnamefont
  {Steinhardt}},\ }\href {\doibase 10.1086/306436} {\bibfield  {journal}
  {\bibinfo  {journal} {Astrophys. J.}\ }\textbf {\bibinfo {volume} {508}},\
  \bibinfo {pages} {483} (\bibinfo {year} {1998})},\ \Eprint
  {http://arxiv.org/abs/astro-ph/9804015} {arXiv:astro-ph/9804015 [astro-ph]}
  \BibitemShut {NoStop}%
\bibitem [{\citenamefont {Ade}\ \emph {et~al.}(2016{\natexlab{b}})\citenamefont
  {Ade} \emph {et~al.}}]{Ade:2015XIV}%
  \BibitemOpen
  \bibfield  {author} {\bibinfo {author} {\bibfnamefont {P.~A.~R.}\
  \bibnamefont {Ade}} \emph {et~al.} (\bibinfo {collaboration} {Planck}),\
  }\href {\doibase 10.1051/0004-6361/201525814} {\bibfield  {journal} {\bibinfo
   {journal} {Astron. Astrophys.}\ }\textbf {\bibinfo {volume} {594}},\
  \bibinfo {pages} {A14} (\bibinfo {year} {2016}{\natexlab{b}})},\ \Eprint
  {http://arxiv.org/abs/1502.01590} {arXiv:1502.01590 [astro-ph.CO]}
  \BibitemShut {NoStop}%
\bibitem [{\citenamefont {Efstathiou}\ and\ \citenamefont
  {Bond}(1999)}]{Efstathiou:1998xx}%
  \BibitemOpen
  \bibfield  {author} {\bibinfo {author} {\bibfnamefont {G.}~\bibnamefont
  {Efstathiou}}\ and\ \bibinfo {author} {\bibfnamefont {J.~R.}\ \bibnamefont
  {Bond}},\ }\href {\doibase 10.1046/j.1365-8711.1999.02274.x} {\bibfield
  {journal} {\bibinfo  {journal} {Mon. Not. Roy. Astron. Soc.}\ }\textbf
  {\bibinfo {volume} {304}},\ \bibinfo {pages} {75} (\bibinfo {year} {1999})},\
  \Eprint {http://arxiv.org/abs/astro-ph/9807103} {arXiv:astro-ph/9807103
  [astro-ph]} \BibitemShut {NoStop}%
\bibitem [{\citenamefont {Hu}\ and\ \citenamefont
  {Sugiyama}(1996)}]{Hu:1995en}%
  \BibitemOpen
  \bibfield  {author} {\bibinfo {author} {\bibfnamefont {W.}~\bibnamefont
  {Hu}}\ and\ \bibinfo {author} {\bibfnamefont {N.}~\bibnamefont {Sugiyama}},\
  }\href {\doibase 10.1086/177989} {\bibfield  {journal} {\bibinfo  {journal}
  {Astrophys. J.}\ }\textbf {\bibinfo {volume} {471}},\ \bibinfo {pages} {542}
  (\bibinfo {year} {1996})},\ \Eprint {http://arxiv.org/abs/astro-ph/9510117}
  {arXiv:astro-ph/9510117 [astro-ph]} \BibitemShut {NoStop}%
\bibitem [{\citenamefont {Beutler}\ \emph {et~al.}(2011)\citenamefont
  {Beutler}, \citenamefont {Blake}, \citenamefont {Colless}, \citenamefont
  {Jones}, \citenamefont {Staveley-Smith}, \citenamefont {Campbell},
  \citenamefont {Parker}, \citenamefont {Saunders},\ and\ \citenamefont
  {Watson}}]{Beutler:2011hx}%
  \BibitemOpen
  \bibfield  {author} {\bibinfo {author} {\bibfnamefont {F.}~\bibnamefont
  {Beutler}}, \bibinfo {author} {\bibfnamefont {C.}~\bibnamefont {Blake}},
  \bibinfo {author} {\bibfnamefont {M.}~\bibnamefont {Colless}}, \bibinfo
  {author} {\bibfnamefont {D.~H.}\ \bibnamefont {Jones}}, \bibinfo {author}
  {\bibfnamefont {L.}~\bibnamefont {Staveley-Smith}}, \bibinfo {author}
  {\bibfnamefont {L.}~\bibnamefont {Campbell}}, \bibinfo {author}
  {\bibfnamefont {Q.}~\bibnamefont {Parker}}, \bibinfo {author} {\bibfnamefont
  {W.}~\bibnamefont {Saunders}}, \ and\ \bibinfo {author} {\bibfnamefont
  {F.}~\bibnamefont {Watson}},\ }\href {\doibase
  10.1111/j.1365-2966.2011.19250.x} {\bibfield  {journal} {\bibinfo  {journal}
  {Mon. Not. Roy. Astron. Soc.}\ }\textbf {\bibinfo {volume} {416}},\ \bibinfo
  {pages} {3017} (\bibinfo {year} {2011})},\ \Eprint
  {http://arxiv.org/abs/1106.3366} {arXiv:1106.3366 [astro-ph.CO]} \BibitemShut
  {NoStop}%
\bibitem [{\citenamefont {Padmanabhan}\ \emph {et~al.}(2012)\citenamefont
  {Padmanabhan}, \citenamefont {Xu}, \citenamefont {Eisenstein}, \citenamefont
  {Scalzo}, \citenamefont {Cuesta}, \citenamefont {Mehta},\ and\ \citenamefont
  {Kazin}}]{Padmanabhan:2012hf}%
  \BibitemOpen
  \bibfield  {author} {\bibinfo {author} {\bibfnamefont {N.}~\bibnamefont
  {Padmanabhan}}, \bibinfo {author} {\bibfnamefont {X.}~\bibnamefont {Xu}},
  \bibinfo {author} {\bibfnamefont {D.~J.}\ \bibnamefont {Eisenstein}},
  \bibinfo {author} {\bibfnamefont {R.}~\bibnamefont {Scalzo}}, \bibinfo
  {author} {\bibfnamefont {A.~J.}\ \bibnamefont {Cuesta}}, \bibinfo {author}
  {\bibfnamefont {K.~T.}\ \bibnamefont {Mehta}}, \ and\ \bibinfo {author}
  {\bibfnamefont {E.}~\bibnamefont {Kazin}},\ }\href {\doibase
  10.1111/j.1365-2966.2012.21888.x} {\bibfield  {journal} {\bibinfo  {journal}
  {Mon. Not. Roy. Astron. Soc.}\ }\textbf {\bibinfo {volume} {427}},\ \bibinfo
  {pages} {2132} (\bibinfo {year} {2012})},\ \Eprint
  {http://arxiv.org/abs/1202.0090} {arXiv:1202.0090 [astro-ph.CO]} \BibitemShut
  {NoStop}%
\bibitem [{\citenamefont {Ross}\ \emph {et~al.}(2015)\citenamefont {Ross},
  \citenamefont {Samushia}, \citenamefont {Howlett}, \citenamefont {Percival},
  \citenamefont {Burden},\ and\ \citenamefont {Manera}}]{Ross:2014qpa}%
  \BibitemOpen
  \bibfield  {author} {\bibinfo {author} {\bibfnamefont {A.~J.}\ \bibnamefont
  {Ross}}, \bibinfo {author} {\bibfnamefont {L.}~\bibnamefont {Samushia}},
  \bibinfo {author} {\bibfnamefont {C.}~\bibnamefont {Howlett}}, \bibinfo
  {author} {\bibfnamefont {W.~J.}\ \bibnamefont {Percival}}, \bibinfo {author}
  {\bibfnamefont {A.}~\bibnamefont {Burden}}, \ and\ \bibinfo {author}
  {\bibfnamefont {M.}~\bibnamefont {Manera}},\ }\href {\doibase
  10.1093/mnras/stv154} {\bibfield  {journal} {\bibinfo  {journal} {Mon. Not.
  Roy. Astron. Soc.}\ }\textbf {\bibinfo {volume} {449}},\ \bibinfo {pages}
  {835} (\bibinfo {year} {2015})},\ \Eprint {http://arxiv.org/abs/1409.3242}
  {arXiv:1409.3242 [astro-ph.CO]} \BibitemShut {NoStop}%
\bibitem [{\citenamefont {Anderson}\ \emph {et~al.}(2014)\citenamefont
  {Anderson} \emph {et~al.}}]{Anderson:2013zyy}%
  \BibitemOpen
  \bibfield  {author} {\bibinfo {author} {\bibfnamefont {L.}~\bibnamefont
  {Anderson}} \emph {et~al.} (\bibinfo {collaboration} {BOSS}),\ }\href
  {\doibase 10.1093/mnras/stu523} {\bibfield  {journal} {\bibinfo  {journal}
  {Mon. Not. Roy. Astron. Soc.}\ }\textbf {\bibinfo {volume} {441}},\ \bibinfo
  {pages} {24} (\bibinfo {year} {2014})},\ \Eprint
  {http://arxiv.org/abs/1312.4877} {arXiv:1312.4877 [astro-ph.CO]} \BibitemShut
  {NoStop}%
\bibitem [{\citenamefont {Kazin}\ \emph {et~al.}(2014)\citenamefont {Kazin}
  \emph {et~al.}}]{Kazin:2014qga}%
  \BibitemOpen
  \bibfield  {author} {\bibinfo {author} {\bibfnamefont {E.~A.}\ \bibnamefont
  {Kazin}} \emph {et~al.},\ }\href {\doibase 10.1093/mnras/stu778} {\bibfield
  {journal} {\bibinfo  {journal} {Mon. Not. Roy. Astron. Soc.}\ }\textbf
  {\bibinfo {volume} {441}},\ \bibinfo {pages} {3524} (\bibinfo {year}
  {2014})},\ \Eprint {http://arxiv.org/abs/1401.0358} {arXiv:1401.0358
  [astro-ph.CO]} \BibitemShut {NoStop}%
\bibitem [{\citenamefont {Alam}\ \emph {et~al.}(2016)\citenamefont {Alam} \emph
  {et~al.}}]{Alam:2016hwk}%
  \BibitemOpen
  \bibfield  {author} {\bibinfo {author} {\bibfnamefont {S.}~\bibnamefont
  {Alam}} \emph {et~al.} (\bibinfo {collaboration} {BOSS}),\ }\href@noop {}
  {\bibfield  {journal} {\bibinfo  {journal} {Submitted to: Mon. Not. Roy.
  Astron. Soc.}\ } (\bibinfo {year} {2016})},\ \Eprint
  {http://arxiv.org/abs/1607.03155} {arXiv:1607.03155 [astro-ph.CO]}
  \BibitemShut {NoStop}%
\bibitem [{\citenamefont {Eisenstein}\ and\ \citenamefont
  {Hu}(1998)}]{Eisenstein:1997ik}%
  \BibitemOpen
  \bibfield  {author} {\bibinfo {author} {\bibfnamefont {D.~J.}\ \bibnamefont
  {Eisenstein}}\ and\ \bibinfo {author} {\bibfnamefont {W.}~\bibnamefont
  {Hu}},\ }\href {\doibase 10.1086/305424} {\bibfield  {journal} {\bibinfo
  {journal} {Astrophys. J.}\ }\textbf {\bibinfo {volume} {496}},\ \bibinfo
  {pages} {605} (\bibinfo {year} {1998})},\ \Eprint
  {http://arxiv.org/abs/astro-ph/9709112} {arXiv:astro-ph/9709112 [astro-ph]}
  \BibitemShut {NoStop}%
\bibitem [{\citenamefont {Song}\ and\ \citenamefont
  {Percival}(2009)}]{Song:2008qt}%
  \BibitemOpen
  \bibfield  {author} {\bibinfo {author} {\bibfnamefont {Y.-S.}\ \bibnamefont
  {Song}}\ and\ \bibinfo {author} {\bibfnamefont {W.~J.}\ \bibnamefont
  {Percival}},\ }\href {\doibase 10.1088/1475-7516/2009/10/004} {\bibfield
  {journal} {\bibinfo  {journal} {JCAP}\ }\textbf {\bibinfo {volume} {0910}},\
  \bibinfo {pages} {004} (\bibinfo {year} {2009})},\ \Eprint
  {http://arxiv.org/abs/0807.0810} {arXiv:0807.0810 [astro-ph]} \BibitemShut
  {NoStop}%
\bibitem [{\citenamefont {Kazantzidis}\ and\ \citenamefont
  {Perivolaropoulos}(2018)}]{Kazantzidis:2018rnb}%
  \BibitemOpen
  \bibfield  {author} {\bibinfo {author} {\bibfnamefont {L.}~\bibnamefont
  {Kazantzidis}}\ and\ \bibinfo {author} {\bibfnamefont {L.}~\bibnamefont
  {Perivolaropoulos}},\ }\href {\doibase 10.1103/PhysRevD.97.103503} {\bibfield
   {journal} {\bibinfo  {journal} {Phys. Rev.}\ }\textbf {\bibinfo {volume}
  {D97}},\ \bibinfo {pages} {103503} (\bibinfo {year} {2018})},\ \Eprint
  {http://arxiv.org/abs/1803.01337} {arXiv:1803.01337 [astro-ph.CO]}
  \BibitemShut {NoStop}%
\bibitem [{\citenamefont {Macaulay}\ \emph {et~al.}(2013)\citenamefont
  {Macaulay}, \citenamefont {Wehus},\ and\ \citenamefont
  {Eriksen}}]{Macaulay:2013swa}%
  \BibitemOpen
  \bibfield  {author} {\bibinfo {author} {\bibfnamefont {E.}~\bibnamefont
  {Macaulay}}, \bibinfo {author} {\bibfnamefont {I.~K.}\ \bibnamefont {Wehus}},
  \ and\ \bibinfo {author} {\bibfnamefont {H.~K.}\ \bibnamefont {Eriksen}},\
  }\href {\doibase 10.1103/PhysRevLett.111.161301} {\bibfield  {journal}
  {\bibinfo  {journal} {Phys. Rev. Lett.}\ }\textbf {\bibinfo {volume} {111}},\
  \bibinfo {pages} {161301} (\bibinfo {year} {2013})},\ \Eprint
  {http://arxiv.org/abs/1303.6583} {arXiv:1303.6583 [astro-ph.CO]} \BibitemShut
  {NoStop}%
\bibitem [{\citenamefont {Taddei}\ and\ \citenamefont
  {Amendola}(2015)}]{Taddei:2014wqa}%
  \BibitemOpen
  \bibfield  {author} {\bibinfo {author} {\bibfnamefont {L.}~\bibnamefont
  {Taddei}}\ and\ \bibinfo {author} {\bibfnamefont {L.}~\bibnamefont
  {Amendola}},\ }\href {\doibase 10.1088/1475-7516/2015/02/001} {\bibfield
  {journal} {\bibinfo  {journal} {JCAP}\ }\textbf {\bibinfo {volume} {1502}},\
  \bibinfo {pages} {001} (\bibinfo {year} {2015})},\ \Eprint
  {http://arxiv.org/abs/1408.3520} {arXiv:1408.3520 [astro-ph.CO]} \BibitemShut
  {NoStop}%
\bibitem [{\citenamefont {Taddei}\ \emph {et~al.}(2016)\citenamefont {Taddei},
  \citenamefont {Martinelli},\ and\ \citenamefont {Amendola}}]{Taddei:2016iku}%
  \BibitemOpen
  \bibfield  {author} {\bibinfo {author} {\bibfnamefont {L.}~\bibnamefont
  {Taddei}}, \bibinfo {author} {\bibfnamefont {M.}~\bibnamefont {Martinelli}},
  \ and\ \bibinfo {author} {\bibfnamefont {L.}~\bibnamefont {Amendola}},\
  }\href {\doibase 10.1088/1475-7516/2016/12/032} {\bibfield  {journal}
  {\bibinfo  {journal} {JCAP}\ }\textbf {\bibinfo {volume} {1612}},\ \bibinfo
  {pages} {032} (\bibinfo {year} {2016})},\ \Eprint
  {http://arxiv.org/abs/1604.01059} {arXiv:1604.01059 [astro-ph.CO]}
  \BibitemShut {NoStop}%
\bibitem [{\citenamefont {Verde}\ \emph {et~al.}(2013)\citenamefont {Verde},
  \citenamefont {Protopapas},\ and\ \citenamefont {Jimenez}}]{Verde:2013wza}%
  \BibitemOpen
  \bibfield  {author} {\bibinfo {author} {\bibfnamefont {L.}~\bibnamefont
  {Verde}}, \bibinfo {author} {\bibfnamefont {P.}~\bibnamefont {Protopapas}}, \
  and\ \bibinfo {author} {\bibfnamefont {R.}~\bibnamefont {Jimenez}},\ }\href
  {\doibase 10.1016/j.dark.2013.09.002} {\bibfield  {journal} {\bibinfo
  {journal} {Phys. Dark Univ.}\ }\textbf {\bibinfo {volume} {2}},\ \bibinfo
  {pages} {166} (\bibinfo {year} {2013})},\ \Eprint
  {http://arxiv.org/abs/1306.6766} {arXiv:1306.6766 [astro-ph.CO]} \BibitemShut
  {NoStop}%
\bibitem [{\citenamefont {Jeffreys}(1961)}]{jeffreys1961theory}%
  \BibitemOpen
  \bibfield  {author} {\bibinfo {author} {\bibfnamefont {H.}~\bibnamefont
  {Jeffreys}},\ }\href {https://books.google.com.br/books?id=AavQAAAAMAAJ}
  {\emph {\bibinfo {title} {Theory of Probability}}},\ The International series
  of monographs on physics\ (\bibinfo  {publisher} {Clarendon Press},\ \bibinfo
  {year} {1961})\BibitemShut {NoStop}%
\bibitem [{\citenamefont {Trotta}(2008)}]{Trotta:2008qt}%
  \BibitemOpen
  \bibfield  {author} {\bibinfo {author} {\bibfnamefont {R.}~\bibnamefont
  {Trotta}},\ }\href {\doibase 10.1080/00107510802066753} {\bibfield  {journal}
  {\bibinfo  {journal} {Contemp. Phys.}\ }\textbf {\bibinfo {volume} {49}},\
  \bibinfo {pages} {71} (\bibinfo {year} {2008})},\ \Eprint
  {http://arxiv.org/abs/0803.4089} {arXiv:0803.4089 [astro-ph]} \BibitemShut
  {NoStop}%
\bibitem [{\citenamefont {Bonilla~Rivera}\ and\ \citenamefont
  {Farieta}(2016)}]{Rivera:2016zzr}%
  \BibitemOpen
  \bibfield  {author} {\bibinfo {author} {\bibfnamefont {A.}~\bibnamefont
  {Bonilla~Rivera}}\ and\ \bibinfo {author} {\bibfnamefont {J.~G.}\
  \bibnamefont {Farieta}},\ }\href@noop {} {\  (\bibinfo {year} {2016})},\
  \Eprint {http://arxiv.org/abs/1605.01984} {arXiv:1605.01984 [astro-ph.CO]}
  \BibitemShut {NoStop}%
\bibitem [{\citenamefont {Burnham}\ and\ \citenamefont
  {Anderson}(2013)}]{burnham2013model}%
  \BibitemOpen
  \bibfield  {author} {\bibinfo {author} {\bibfnamefont {K.}~\bibnamefont
  {Burnham}}\ and\ \bibinfo {author} {\bibfnamefont {D.}~\bibnamefont
  {Anderson}},\ }\href {https://books.google.com.br/books?id=W63hBwAAQBAJ}
  {\emph {\bibinfo {title} {Model Selection and Inference: A Practical
  Information-Theoretic Approach}}}\ (\bibinfo  {publisher} {Springer New
  York},\ \bibinfo {year} {2013})\BibitemShut {NoStop}%
\bibitem [{\citenamefont {Pérez-Romero}\ and\ \citenamefont
  {Nesseris}(2018)}]{Perez-Romero:2017njc}%
  \BibitemOpen
  \bibfield  {author} {\bibinfo {author} {\bibfnamefont {J.}~\bibnamefont
  {Pérez-Romero}}\ and\ \bibinfo {author} {\bibfnamefont {S.}~\bibnamefont
  {Nesseris}},\ }\href {\doibase 10.1103/PhysRevD.97.023525} {\bibfield
  {journal} {\bibinfo  {journal} {Phys. Rev.}\ }\textbf {\bibinfo {volume}
  {D97}},\ \bibinfo {pages} {023525} (\bibinfo {year} {2018})},\ \Eprint
  {http://arxiv.org/abs/1710.05634} {arXiv:1710.05634 [astro-ph.CO]}
  \BibitemShut {NoStop}%
\bibitem [{\citenamefont {Evslin}\ \emph {et~al.}(2017)\citenamefont {Evslin},
  \citenamefont {Sen},\ and\ \citenamefont {Ruchika}}]{Evslin:2017qdn}%
  \BibitemOpen
  \bibfield  {author} {\bibinfo {author} {\bibfnamefont {J.}~\bibnamefont
  {Evslin}}, \bibinfo {author} {\bibfnamefont {A.~A.}\ \bibnamefont {Sen}}, \
  and\ \bibinfo {author} {\bibnamefont {Ruchika}},\ }\href@noop {} {\
  (\bibinfo {year} {2017})},\ \Eprint {http://arxiv.org/abs/1711.01051}
  {arXiv:1711.01051 [astro-ph.CO]} \BibitemShut {NoStop}%
\bibitem [{\citenamefont {Riess}\ \emph {et~al.}(2016)\citenamefont {Riess}
  \emph {et~al.}}]{Riess:2016jrr}%
  \BibitemOpen
  \bibfield  {author} {\bibinfo {author} {\bibfnamefont {A.~G.}\ \bibnamefont
  {Riess}} \emph {et~al.},\ }\href {\doibase 10.3847/0004-637X/826/1/56}
  {\bibfield  {journal} {\bibinfo  {journal} {Astrophys. J.}\ }\textbf
  {\bibinfo {volume} {826}},\ \bibinfo {pages} {56} (\bibinfo {year} {2016})},\
  \Eprint {http://arxiv.org/abs/1604.01424} {arXiv:1604.01424 [astro-ph.CO]}
  \BibitemShut {NoStop}%
\bibitem [{\citenamefont {Tully}\ \emph {et~al.}(2016)\citenamefont {Tully},
  \citenamefont {Courtois},\ and\ \citenamefont {Sorce}}]{Tully:2016ppz}%
  \BibitemOpen
  \bibfield  {author} {\bibinfo {author} {\bibfnamefont {R.~B.}\ \bibnamefont
  {Tully}}, \bibinfo {author} {\bibfnamefont {H.~M.}\ \bibnamefont {Courtois}},
  \ and\ \bibinfo {author} {\bibfnamefont {J.~G.}\ \bibnamefont {Sorce}},\
  }\href {\doibase 10.3847/0004-6256/152/2/50} {\bibfield  {journal} {\bibinfo
  {journal} {Astron. J.}\ }\textbf {\bibinfo {volume} {152}},\ \bibinfo {pages}
  {50} (\bibinfo {year} {2016})},\ \Eprint {http://arxiv.org/abs/1605.01765}
  {arXiv:1605.01765 [astro-ph.CO]} \BibitemShut {NoStop}%
\bibitem [{\citenamefont {Feeney}\ \emph {et~al.}(2018)\citenamefont {Feeney},
  \citenamefont {Peiris}, \citenamefont {Williamson}, \citenamefont {Nissanke},
  \citenamefont {Mortlock}, \citenamefont {Alsing},\ and\ \citenamefont
  {Scolnic}}]{Feeney:2018mkj}%
  \BibitemOpen
  \bibfield  {author} {\bibinfo {author} {\bibfnamefont {S.~M.}\ \bibnamefont
  {Feeney}}, \bibinfo {author} {\bibfnamefont {H.~V.}\ \bibnamefont {Peiris}},
  \bibinfo {author} {\bibfnamefont {A.~R.}\ \bibnamefont {Williamson}},
  \bibinfo {author} {\bibfnamefont {S.~M.}\ \bibnamefont {Nissanke}}, \bibinfo
  {author} {\bibfnamefont {D.~J.}\ \bibnamefont {Mortlock}}, \bibinfo {author}
  {\bibfnamefont {J.}~\bibnamefont {Alsing}}, \ and\ \bibinfo {author}
  {\bibfnamefont {D.}~\bibnamefont {Scolnic}},\ }\href@noop {} {\  (\bibinfo
  {year} {2018})},\ \Eprint {http://arxiv.org/abs/1802.03404} {arXiv:1802.03404
  [astro-ph.CO]} \BibitemShut {NoStop}%
\bibitem [{\citenamefont {da~Silva}\ and\ \citenamefont
  {Cavalcanti}(2018)}]{daSilva:2018sim}%
  \BibitemOpen
  \bibfield  {author} {\bibinfo {author} {\bibfnamefont {G.~P.}\ \bibnamefont
  {da~Silva}}\ and\ \bibinfo {author} {\bibfnamefont {A.~G.}\ \bibnamefont
  {Cavalcanti}},\ }\href@noop {} {\  (\bibinfo {year} {2018})},\ \Eprint
  {http://arxiv.org/abs/1805.06849} {arXiv:1805.06849 [astro-ph.CO]}
  \BibitemShut {NoStop}%
\bibitem [{\citenamefont {Gómez-Valent}\ and\ \citenamefont
  {Amendola}(2018)}]{Gomez-Valent:2018hwc}%
  \BibitemOpen
  \bibfield  {author} {\bibinfo {author} {\bibfnamefont {A.}~\bibnamefont
  {Gómez-Valent}}\ and\ \bibinfo {author} {\bibfnamefont {L.}~\bibnamefont
  {Amendola}},\ }\href {\doibase 10.1088/1475-7516/2018/04/051} {\bibfield
  {journal} {\bibinfo  {journal} {JCAP}\ }\textbf {\bibinfo {volume} {1804}},\
  \bibinfo {pages} {051} (\bibinfo {year} {2018})},\ \Eprint
  {http://arxiv.org/abs/1802.01505} {arXiv:1802.01505 [astro-ph.CO]}
  \BibitemShut {NoStop}%
\end{thebibliography}%

\appendix

\section{\texttt{mBayes}}\label{mBayes}

The results presented in this paper were obtained using \texttt{mBayes}, a numerical package that aims at helping researchers to effortlessly carry out Bayesian inference within Wolfram Mathematica.
The analysis part is completely automatized while the posterior exploration part only needs adjusting the ``glue code'' section.
At the moment, the following features are implemented:
\begin{itemize}

\item multivariate and flat priors,

\item variables can be easily fixed without editing the glue code,

\item Fisher matrix approximation for likelihood and posterior,

\item Fisher and fast numerical evidence,

\item grid optimization with Fisher,

\item optimized parallel computation and exportation,

\item automatized exportation of results with consistent labeling,

\item confidence levels (actual and gaussian),

\item combinations of triangular plots.

\end{itemize}
An MCMC sampler and further optimizations will be implemented in the near future.
\texttt{mBayes} is available at \href{https://github.com/valerio-marra/mBayes}{github.com/valerio-marra/mBayes}.

\section{Triangular plots}\label{triplots}

Here, we show the triangular plots relative to section~\ref{result}. The plots are important to understand correlations and degeneracies between the various parameters.

\begin{figure*}
  \centering
   \includegraphics[width=0.75\textwidth]{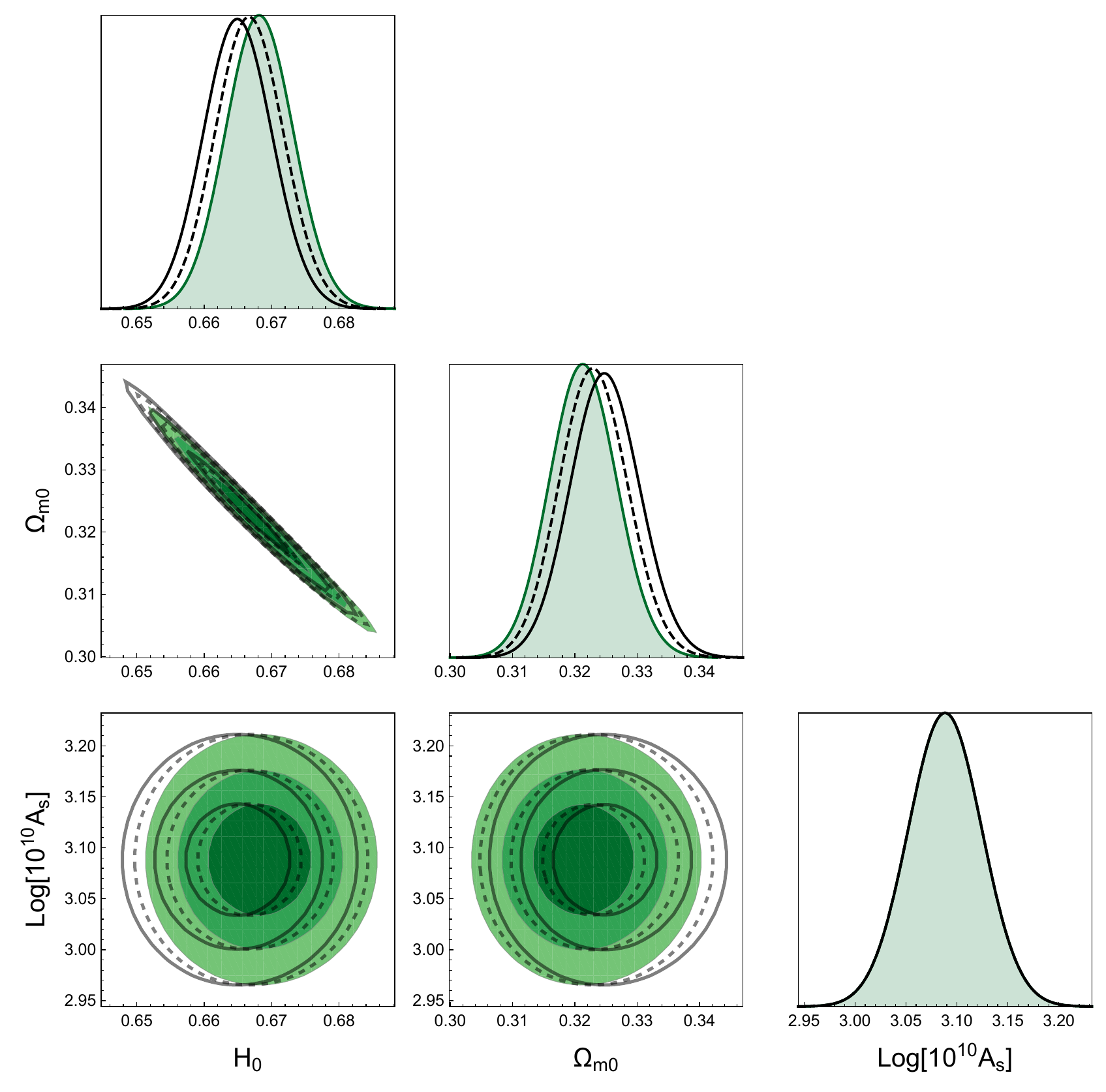}
  \caption{
1-, 2- and 3$\sigma$ marginalized constraints on the parameters of the $\Lambda$CDM model for the likelihoods from equation~\eqref{litot} without cosmic variance (green contours) and with cosmic variance (dashed black contours). See section~\ref{result} for details.}
  \label{fig:dH1_combo_r5}
\end{figure*}
\begin{figure*}
  \centering
   \includegraphics[width=0.95\textwidth]{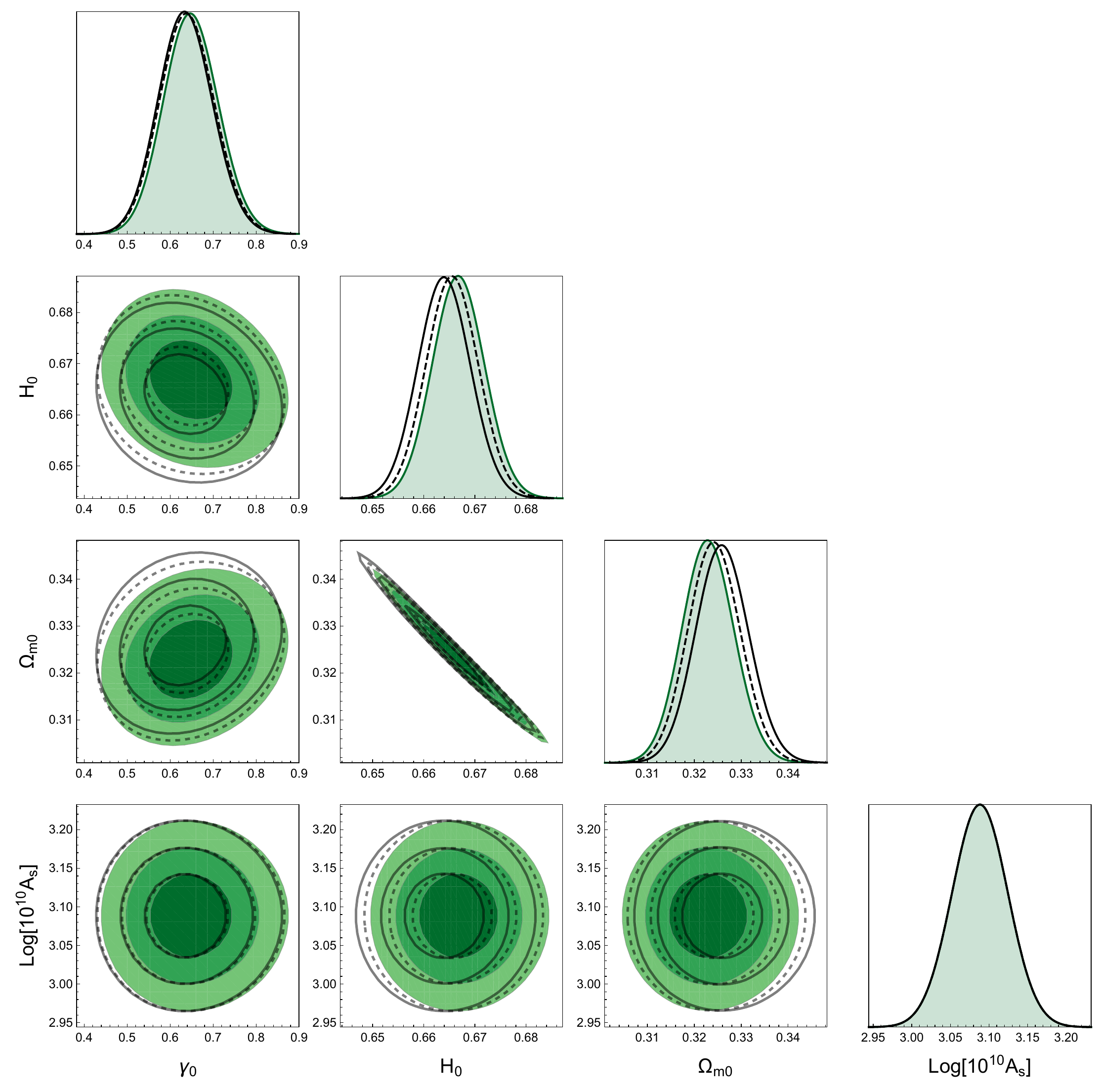}
  \caption{1-, 2- and 3$\sigma$ marginalized constraints on the parameters of the $\gamma$CDM model for the likelihoods from equation~\eqref{litot} without cosmic variance (green contours) and with cosmic variance  (dashed black contours). See section~\ref{result} for details.}
  \label{fig:dH2_combo_r5}
\end{figure*}
\begin{figure*}
  \centering
   \includegraphics[width=0.95\textwidth]{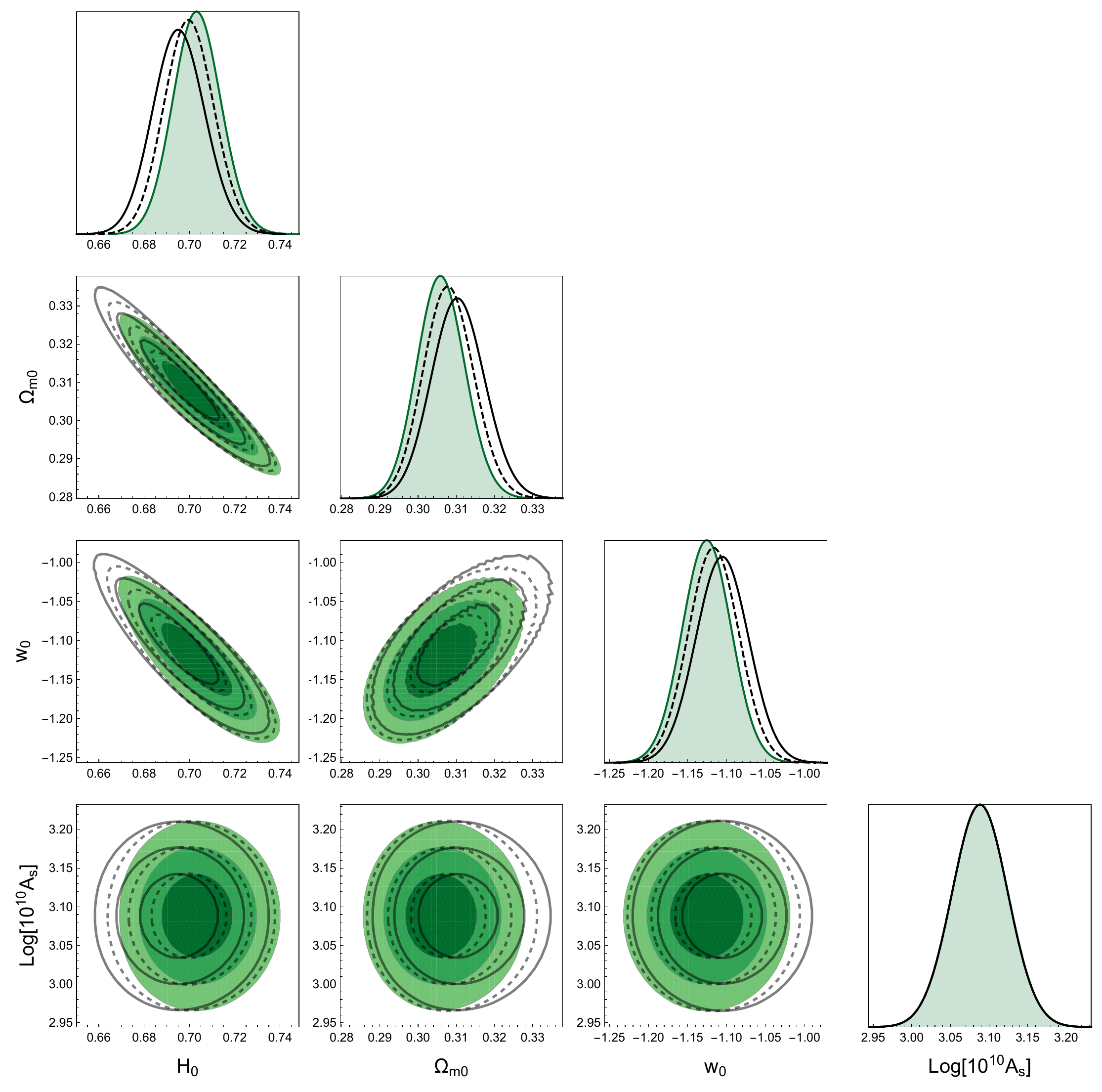}
  \caption{1-, 2- and 3$\sigma$ marginalized constraints on the parameters of the $w$CDM model for the likelihoods from equation~\eqref{litot} without cosmic variance (green contours) and with cosmic variance (dashed black contours). See section~\ref{result} for details.}
  \label{fig:dH7_combo_r5}
\end{figure*}
\begin{figure*}
  \centering
   \includegraphics[width=0.95\textwidth]{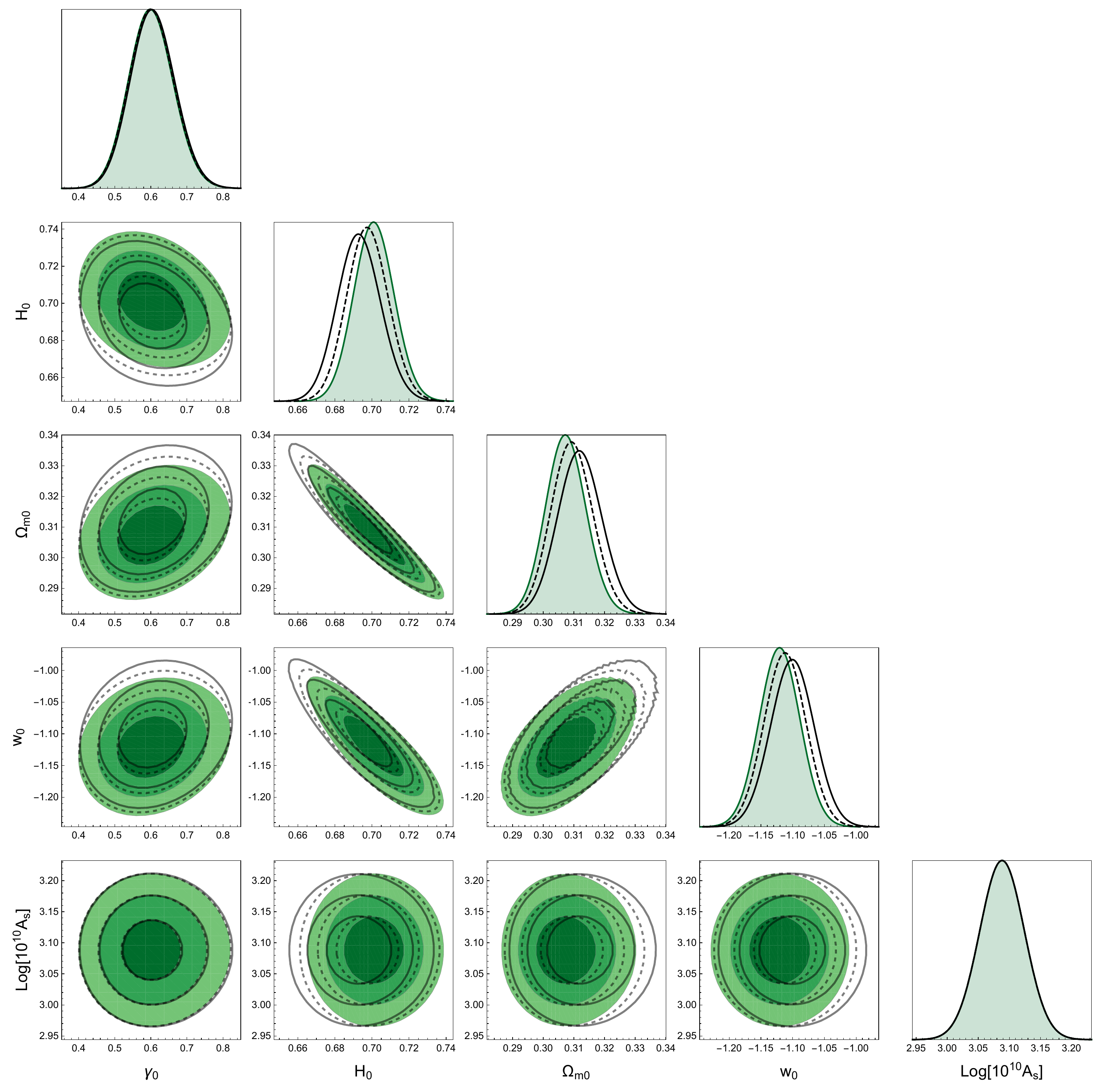}
  \caption{1-, 2- and 3$\sigma$ marginalized constraints on the parameters of the $\gamma w$CDM model for the likelihoods from equation~\eqref{litot} without cosmic variance (green contours) and with cosmic variance (dashed black contours). See section~\ref{result} for details.}
  \label{fig:dH6_combo_r5}
\end{figure*}

\section{Fisher matrices and best-fit parameters}\label{fisher}

\begin{table}[t!]
\centering
\begin{tabular}{|l|c|c|c|c|c|c|}
\hline
\multicolumn{6}{|c|}{\bf Analysis with $\chi^2_{\rm tot,0}$  (without $\sigma_{\rm cv}$)} \\
\hline
Model &  $ \gamma$ & $H_0$[km/s/Mpc] & $\Omega_{m0}$ & $w$ & $\ln{(10^{10} A_s)}$ \\ \hline
$\Lambda$CDM  & - & $66.8$ &  $0.321$ & - & $3.088$ \\ \hline
$\gamma$CDM  & $0.65$ & $66.7$ & $0.323$ & - & $3.088$ \\ \hline
$w$CDM  & - & $70.4$ & $0.306$ & $-1.13$ & $ 3.088$
\\ \hline
$\gamma w$CDM  & $0.60$ & $70.1$ & $0.307$ & $ -1.12$ & $3.088$ \\ \hline
\hline
\multicolumn{6}{|c|}{\bf Analysis with $\chi^2_{\rm tot,1}$  (with $\sigma_{\rm cv,1}$)} \\ \hline
Model &  $ \gamma$ & $H_0$[km/s/Mpc] & $\Omega_{m0}$ & $w$ & $\ln{(10^{10} A_s)}$ \\ \hline
$\Lambda$CDM  & - & $66.7$ & $0.323$ & - & $3.088$ \\ \hline
$\gamma$CDM  & $0.64$ & $66.6$ & $0.324$ & - & $3.088$ \\ \hline
$w$CDM  & - &$69.9$ & $0.308$ & $-1.12$ & $3.088$ \\ \hline
$\gamma w$CDM  & $0.60$ & $69.8$ & $0.309$ & $-1.11$ & $3.088$ \\ \hline
\hline
\multicolumn{6}{|c|}{\bf Analysis with $\chi^2_{\rm tot,2}$  (with $\sigma_{\rm cv,2}$)} \\ \hline
Model &  $ \gamma$ & $H_0$[km/s/Mpc] & $\Omega_{m0}$ & $w$ & $\ln{(10^{10} A_s)}$ \\ \hline
$\Lambda$CDM  & - & $66.5$ & $0.325$ & - & $3.088$ \\ \hline
$\gamma$CDM  & $0.63$ & $66.4$ & $0.326$ & - & $3.088$ \\ \hline
$w$CDM  & - &$69.5$ & $0.310$ & $-1.10$ & $3.088$ \\ \hline
$\gamma w$CDM  & $0.60$ & $69.3$ & $0.312$ & $-1.10$ & $3.088$ \\ \hline
\end{tabular}
\caption{Best-fit parameters.}
\label{tabBFX}
\end{table}

Here, we list the Fisher matrices $L_{\alpha \beta}$ and the best-fit parameters (see Table~\ref{tabBFX}) relative to the likelihoods considered in this work.
Using the latter one can accurately approximate the (normalized) posterior.
The Fisher matrices do not change substantially; this means that cosmic variance mainly shifts the best-fit vector.

\begin{widetext}
\begin{align}
L_0&=10^6
\left(
\begin{array}{ccc}
H_0 & \Omega_{m} & \ln \! 10^{10}\! A_s \\
\hline
 1.7 & 1.5 & 0 \\
 \text{} & 1.4 & 0 \\
 \text{} & \text{} & 0.00077 \\
\end{array}
\right) \,, \nonumber \\
L_{0,\text{cv,1}}&=10^6
\left(
\begin{array}{ccc}
 1.7 & 1.5 & 0 \\
 \text{} & 1.4 & 0 \\
 \text{} & \text{} & 0.00077 \\
\end{array}
\right) \,, \nonumber \\
L_{0,\text{cv,2}}&=10^6
\left(
\begin{array}{ccc}
 1.7 & 1.5 & 0 \\
 \text{} & 1.4 & 0 \\
 \text{} & \text{} & 0.00077 \\
\end{array}
\right) \,, \nonumber \\
L_\gamma&=10^6
\left(
\begin{array}{cccc}
\gamma & H_0 & \Omega_{m} & \ln \! 10^{10}\! A_s \\
\hline
 0.00026 &0 & -0.00056 & 0 \\
 \text{} & 1.7 & 1.5 & 0\\
 \text{} & \text{} & 1.4 & 0 \\
 \text{} & \text{} & \text{} & 0.00077 \\
\end{array}
\right)  \,, \nonumber \\
L_{\gamma,\text{cv,1}}&=10^6
\left(
\begin{array}{cccc}
 0.00026 & -0.000081 & -0.00055 &0 \\
 \text{} & 1.7 & 1.5 & 0 \\
 \text{} & \text{} & 1.4 & 0 \\
 \text{} & \text{} & \text{} & 0.00077 \\
\end{array}
\right) \,, \nonumber \\
L_{\gamma,\text{cv,2}}&=10^6
\left(
\begin{array}{cccc}
 0.00026 & -0.00014 & -0.00055 & 0 \\
 \text{} & 1.7 & 1.5 & 0 \\
 \text{} & \text{} & 1.4 & 0 \\
 \text{} & \text{} & \text{} & 0.00077 \\
\end{array}
\right) \,, \nonumber \\
L_w&=10^6
\left(
\begin{array}{cccc}
H_0 & \Omega_{m} & w &\ln \! 10^{10}\! A_s \\
\hline
 1.3 & 1.5 & 0.2 & 0 \\
 \text{} & 1.7 & 0.22 & 0 \\
 \text{} & \text{} & 0.032 & 0 \\
 \text{} & \text{} & \text{} & 0.00077 \\
\end{array}
\right) \,, \nonumber \\
L_{w,\text{cv,1}}&=10^6
\left(
\begin{array}{cccc}
 1.3 & 1.5 & 0.2 & 0 \\
 \text{} & 1.6 & 0.22 & 0 \\
 \text{} & \text{} & 0.032 & 0 \\
 \text{} & \text{} & \text{} & 0.00077 \\
\end{array}
\right) \,, \nonumber \\
L_{w,\text{cv,2}}&=10^6
\left(
\begin{array}{cccc}
 1.4 & 1.5 & 0.21 & 0 \\
 \text{} & 1.6 & 0.22 & 0 \\
 \text{} & \text{} & 0.033 &0 \\
 \text{} & \text{} & \text{} & 0.00077 \\
\end{array}
\right) \,, \nonumber \\
L_{\gamma w}&=10^6
\left(
\begin{array}{ccccc}
\gamma & H_0 & \Omega_{m} & w & \ln \! 10^{10}\! A_s \\
\hline
  0.00028 & 0 & -0.00059 & -0.000031 &0 \\
 \text{} & 1.3 & 1.5 & 0.2 & 0 \\
 \text{} & \text{} & 1.7 & 0.22 & 0 \\
 \text{} & \text{} & \text{} & 0.032 & 0 \\
 \text{} & \text{} & \text{} & \text{} & 0.00077 \\
\end{array}
\right) \,, \nonumber \\
L_{\gamma w,\text{cv,1}}&=10^6
\left(
\begin{array}{ccccc}
 0.00028 & -0.000052 & -0.00058 & -0.000032 & 0 \\
 \text{} & 1.3 & 1.5 & 0.2 & 0 \\
 \text{} & \text{} & 1.6 & 0.22 & 0 \\
 \text{} & \text{} & \text{} & 0.032 & 0 \\
 \text{} & \text{} & \text{} & \text{} & 0.00077 \\
\end{array}
\right) \,, \nonumber \\
L_{\gamma w,\text{cv,2}}&=10^6
\left(
\begin{array}{ccccc}
 0.00028 & -0.000088 & -0.00058 & -0.000032 &0 \\
 \text{} & 1.4 & 1.5 & 0.21 & 0 \\
 \text{} & \text{} & 1.6 & 0.22 & 0 \\
 \text{} & \text{} & \text{} & 0.033 & 0 \\
 \text{} & \text{} & \text{} & \text{} & 0.00077 \\
\end{array}
\right) \,. \nonumber 
\end{align}
\end{widetext}

\end{document}